\pdfoutput=1
\documentclass[10pt]{article}

\usepackage{algorithm}
\usepackage{color}
\usepackage{bbm}
\usepackage{epsfig}
\usepackage{amssymb,amsmath,amscd, mdwmath}
\usepackage{graphicx,cite,url}
\usepackage{algorithmic}
\usepackage{float}
\usepackage{amsfonts}
\usepackage{syntonly}
\usepackage{multirow}
\usepackage{graphicx}
\usepackage{subfig}
\usepackage{tabularx}
\usepackage{setspace}
\usepackage{stfloats}
\usepackage{amsthm}
\usepackage[labelfont=bf,labelsep=period,justification=raggedright]{caption}

\makeatletter
\renewcommand{\@biblabel}[1]{\quad#1.}
\makeatother

\date{}

\begin{document}

% Title must be 150 characters or less
\begin{flushleft}
{\Large
\textbf{Predicting the number of coauthors       for researchers: A learning  model
 }
%\textbf{Modelling  tipping-point phenomena of scientific coauthorship networks     and citation networks}
}
% Insert Author names, affiliations and corresponding author email.
\\
Zheng Xie$^{1,2, \sharp }$
\\
\bf{1}   College of Liberal Arts and Sciences,   National University of Defense Technology, Changsha,   China. \\
\bf{2} Department of
Mathematics, University of California, Los Angeles,   USA
\\  $^\sharp$ xiezheng81@nudt.edu.cn
 \end{flushleft}
%Sint-Andriesstraat 2, 2000
% Please keep the abstract between 250 and 300 words
\section*{Abstract}
Predicting the   number of coauthors for researchers contributes to understanding the development of team science. However, it is an elusive task due to diversity in the collaboration patterns of researchers. This study provides a learning model for the dynamics of   this variable; the parameters are learned from  empirical data that consist of the number of publications and the number of coauthors at given time intervals. The model is based on   relationship between the annual number of new coauthors   and time given an annual number of publications, the relationship between the annual number of publications and time given a historical number of publications, and   Lotka's law. The assumptions of the model are validated by applying it on the high-quality dblp dataset. The effectiveness of the model is tested on the dataset by   satisfactory  fittings on the evolutionary trend of the number of   coauthors for researchers, the   distribution of this variable, and the occurrence probability of collaboration events. Due to its regression nature, the model has the potential to be extended to assess  the confidence level of the prediction results  and thus has applicability to other empirical research.

\noindent {\bf Keywords:} Coauthorship,  Publication productivity, Data modelling.

\section*{Introduction}

A growing trend of collaboration has  emerged in    current scientific research.
 This trend is  reflected in   increasingly active coauthorship among researchers as solitary authorship diminishes in prevalence\cite{Committee}.
 Coauthorship has  attracted much attention, with analyses of perspectives ranging from contribution\cite{Correa2017,Lu-Zhang2019}, population\cite{Li-Miao2015},        discipline\cite{Moody2004,Wagner2005,XieLL}, country\cite{Perc,Katz1994}, and multination\cite{Russell1995,Glanzel-Schubert1999,Leclerc-Gagne1994,Gomez-Fernandez1999}, to the
  connection with citations\cite{Narin-Stevens1991,Khor-Yu2016}. 
 The emerging field known as  team science draws on diverse disciplinary perspectives to understand  the processes and outcomes of scientific collaboration. Team work has been shown to have a large citation impact\cite{Wuchty-Jones2007,Valderas2007}, transdisciplinary outcomes\cite{Vogel2014}, and high publication productivity\cite{Hall2012}. Uzzi et al indicated that publications with three or more authors showed an increased frequency of   ``tail novelty" (which is a publication's 10th-percentile $z$ score for its journal pairings) over the solo-author rate\cite{Uzzi2013}. They used regression methods to analyze the relationship between the number of citations of a publication and the number of its authors, and found that publications produced by larger teams were associated with a higher citation impact. Wu et al found that the character of publications produced by large teams differs from that of small teams in terms of  development versus disruption\cite{Wu-Wang2019}.  The number of coauthors  is related to the team size. For example,    more than 70\% of researchers in the  empirical  dataset considered here   belong to one team  (see Appendix A).
  Therefore, the prediction has the potential to propose   auxiliary measures for teams' innovation,   impact, and research character.

%Therefore, the prediction results would contribute to team science.

A previous study showed that the assembly mechanisms of
a research team  determine the  structure of coauthorship networks\cite{Newman2004patterns,Guimera-Uzzi2005}.
Much attention has been paid to these networks, and research has been concentrated on   coauthor distribution\cite{BarabasiJeong2002,XieO2016,XieX2017}, followed by  structure\cite{Newman2,Newman3,Newman1}, clustering\cite{Newman2001clust},  homophily\cite{ Newman2002assort, Zhang-Bu2018}, and   applications, e.g., name disambiguation\cite{Amancio2012}.
Researchers  have established a range of models, from  preference attachment to cooperative game theory\cite{Santos,PercM2014,Xie2018Complexity}, to  explore possible  mechanisms for the evolution of the  networks
created by coauthorship.
Most of these models generate a constant number of links for each new   node, which is far from the reality.
To simulate coauthorship networks at full scale, we  need  to know the extent to which researchers collaborate.
Therefore,
  a method of   predicting    the number of  coauthors is   needed.

     %helps to elucidate  to what extent researchers collaborate, and whether the extent changes over time.

 %, particularly  for the multimodality of coauthor distributions.

 Researchers have explored possible factors that increase or decrease the number of new coauthors, such as the institutional prestige\cite{Hunter-Leahey2008}, self-organization\cite{{Melin2000}}, geography\cite{Hoekman-Frenken2010}, discipline or interdiscipline\cite{Rijnsoever-Hessels2011},   and academic  reputation   of researchers\cite{QiMZengA2017}.
Knowledge of the correlated factors helps to predict  the number of coauthors for a given researcher. However, factor analysis in social systems cannot exhaust all possible factors, as the considered factors would be correlated. For example, prestigious institutions that  possess  famous researchers can attract   researchers to collaborate, which in part leads to   multi-university collaborations\cite{Jones-Wuchty2008}. Some have of the identified  correlations between the considered factors and response variables may be caused by unconsidered factors or by the correlations between the considered factors, which are called spurious correlations. Therefore, analyzing factors individually is not recommended in statistical analysis. Accordingly,     a multivariate statistical model to predict  the number of coauthors is   needed.

To
 choose suitable  statistical models, we need to know the detailed features of the distributions of response variables and  the   mechanisms thereof.
Coauthors appear in the process of producing publications;  thus, there is a   need to    predict    publication productivity.
 The number of publications of a researcher can  be  explained by an inhomogeneous Poisson process\cite{Xie2019Scientometrics}.
 A model of piecewise Poisson regression has been proposed  to predict the number of publications\cite{Xie2019ppp0}. Its limitation regarding   the prediction for highly  productive researchers is solved by utilizing  Lotka's law\cite{Xie2019ppp}. Based on the high-quality dblp dataset\footnote[1]{The dblp computer science bibliography proposes a high-quality dataset  that consists of   open bibliographic information on the major journals and conference proceedings of computer science.
 It has been corrected by several methods of name disambiguation, and there are now more than 60,000 manually
confirmed external identities linked with dblp author bibliographies. These confirmed identities guarantee the quality of the dataset. See https://www.dblp.org.}, the effectiveness of these models has been tested by   satisfactory  fittings on the
distribution and the evolutionary trend
of  the number of publications for researchers,  as well as  the occurrence probability of  publication  events.

%  relationship between the annual publication productivity of researchers and their historical publication quantity

This study proposes a learning model to utilize multiple factors to predict the number of coauthors for researchers. Three factors are used as a beginning, namely, time, the historical number of publications, and the historical number of coauthors.
The piecewise Poisson regression   on the training datasets extracted from the dblp dataset,  given an annual number of publication, shows a significant correlation between the annual number of new coauthors of a researcher and time.
However, the   annual number of new coauthors does not follow a Poisson distribution.
Therefore, using the piecewise Poisson regression
and the predicted  annual number of publications   can  provide only
    preliminary   results for the
number of  coauthors.
  The results are modified by
   the  formulae   that address the cumulative advantage of attracting coauthors in terms of the historical  number of coauthors.
  The hyperparameters of   the  formulae are determined  by a genetic algorithm for a good fit  to   validation
datasets.
 The effectiveness of the model is displayed by  a good fit   to the test datasets in terms of  the evolutionary trend of
 the number of coauthors, the  distribution of this variable, and the  occurrence probability of collaboration events.

This paper is organized as follows.  The model and its motivation are described in Sections 2, 3. The empirical  data and experiments are described in Section  4. The results are discussed and conclusions drawn in Section 5.

\section*{Motivation}

\subsection*{The relationship among time,  the number of publications  and coauthors}
A positive correlation between the number of publications of a researcher and the number of his or her coauthors
has been found  in several empirical datasets\cite{Xie2019Scientometrics}, and is  also found in the dataset analyzed   here~(see Appendix A). Correlation does  not indicate    causality, and  arguments
 exist on   whether scientific  collaboration has a positive effect on publishing productivity.
Lee et al found that the number of coauthors is not a significant predictor of the number of  publications\cite{Lee2005}. However,
  Ductor showed that after controlling for endogenous coauthorship formation, unobservable heterogeneity, and time, the effect of intellectual collaboration on the number of an individual's publications becomes positive\cite{DuctorL2015}.
   Therefore, our model does not include the correlation.

% Wardil et al analyzed   the papers of physical review journals to show that authors publishing more often as first authors have
%fewer publications in the short-term than authors publishing more often as last authors\cite{Wardil-Hauert2015}.

%

%The correlation combined with the prediction of publication quantity seems to work for the prediction of  coauthor quantity.

%To predict, we need to know which covariates significantly correlate to $k(t_i)$.

The analysis on the dblp dataset shows that  given an   annual number of publications,   the annual  number    of a researcher's new coauthors   significantly correlates  to    time.
Therefore,
the annual number  of new coauthors
 can be predicted when  the  future annual number of publications    is known.
A previous  model   can  predict the latter variable\cite{Xie2019ppp}, which makes it possible to predict the former variable.
Note that the analysis on the dblp dataset shows that the annual  number of new coauthors does not significantly correlate  to time when considering  all individuals or   individuals with the same historical number of publications.
Therefore,  the
 annual number of publications is  utilized  in our model  as  a middle variable.

 \subsection*{The  distribution of the number of coauthors}

% In addition, the model in Reference\cite{xie_ppp_2} is based on
%the  Lotka's law. Hence, this law is also a component  of this study's foundation.

% In the parlance of latent variable modeling,
%observed (or manifest) variables are those variables
%in the model for which direct, observable scores are
%available.

%\subsection*{From data perspective}
In our study, the coauthor distribution  of a group of researchers   refers to the  distribution of the number of a researcher's coauthors.
To choose a suitable regression model, we need  to know  the distribution features  of the response variables and the mechanisms that generate these features.
 The number of coauthors of  a researcher, as a response variable,  is in part
dependent on his or her number of publications.
 Previous studies on several empirical datasets have shown that
the   distribution  of the number of a researcher's publications
is characterized   by a trichotomy, comprising a generalized Poisson head, a power-law midsection, and an exponential cutoff\cite{Xie2019Scientometrics}.
 The trichotomy  can be derived from a range of ``coin-flipping" behaviors, in which  the probability of observing ``heads" is dependent on   events already observed~\cite{Consul}.

 The event of producing  a publication can be regarded as an analogy of observing ``heads". The probability of publishing is also affected by previous events, and research experiences accumulated in the process of producing publications.
  This is a cumulative advantage that also exists in the analyzed dblp dataset (see Appendix
A).
  It is displayed as a transition from the generated Poisson head to the power-law midsection. The aging of researchers' creativity operates against the cumulative advantage,  and is displayed as the transition from the power-law midsection to the exponential cutoff.

 Lotka's law applies to in empirical datasets\cite{Lotka1926}; that is,
many researchers   have only  one publication.
Meanwhile, the number of  authors of a publication    mainly  follows a generalized  Poisson distribution\cite{Xie2018jasisit}, thus
  inducing the  generalized Poisson parts  of  coauthor distributions.
An increase in   the number of publications  is associated   with an increase in  coauthors, which  induces   the power-law midsection and exponential cutoff  of coauthor distributions.
The analysis reveals  the inhomogeneous  Poisson character of coauthors appearing, which is a key point of our prediction model.

%The analysis shows that the annual new coauthor quantity is generated by
% a   composition of inhomogeneous   Poisson  processes.

%) and the annual publication quantity

 The annual  number   of a researcher's new coauthors    also  depends on his or her historical number of coauthors, which is a cumulative advantage that exists  in the analyzed dataset (see Appendix A).
The effect  of the cumulative advantage   would be nonsignificant  over a short time interval, e.g.,   a year.
At each year $y$,   researchers are partitioned into  $M\times W$ subsets, where       subset $(m,w)$ contains the researchers with  $m$ publications at $y$ and $w$ historical coauthors before $y$.
This partition diminishes the diversity of researchers in terms of their historical number of coauthors and annual number of publications, and  can reveal  the Poisson character of coauthor distributions (see Section 4).
  However,  some of these subsets
are  too   small to use regression. Therefore, this study proposed  a method   to deal the inhomogeneous    Poisson  process
of coauthors appearing for the relatively large subsets that consist  of   researchers with the same  annual number of publications.

\section*{The    model}

\subsection*{Model    terms}

%Partition   the researchers with no more than $I$ publications  into  $I$ subsets according
% to their historical   publication quantity at a time interval  $[T_0,T_1]$, where $I$ is the largest publication quantity  that we can predict.
%For each subset, we estimated its publication  creativity    at a time interval $[T_1,T_2]$.

Consider  the researchers
 who produced publications  at  two intervals $[T_0,T_1]$ and $[T_1,T_2]$.
 Partition  the second one into $J$ intervals with cutpoints $T_1 = t_0 < t_1 <
\cdots < t_J = T_2$. The half-closed interval  $(t_{j-1}, t_{j}]$ is referred to as  the $j$-th time interval, where $j=1,2,...,J$.
  Consider the researchers who produced $m$ publications  at the  $j$-th time interval.
   Let $\xi_{mj}$ be the average number of
 new  coauthors
of these researchers at  the $j$-th time interval.
Let  $\zeta_{mj}$  be the new  coauthor number of each of these researchers   at  the $j$-th time interval.
%It can be regarded as
% the expected value of the distribution  of
%the number of new coauthors obtained by
%such a researcher   at  the $j$-th time interval.

%,  termed
 % the ``publication productivity" of these researchers  at the $j$-th time interval

  Consider the researchers who produced $i$ publications  at the time interval $[T_0,t_{j-1}]$.
Let $\eta_{ij}$ be the average number of these researchers' publications produced    at  the $j$-th time interval.
  Let   $\lambda_{ij}$  be  the number of   publications of each of these researchers   at  the $j$-th time interval.

%, termed  the ``publication creativity" of these researchers  at the $j$-th time interval.

%Let $u$ is random variable of the Poisson distribution of the expected value $\lambda_{ij}$, which is the
%predicted publication quantity of $i$-th subset and $j$-th time interval.
%Let $v$ is random variable of the Poisson distribution of the expected value $\zeta_{uj}$.

A training dataset  is used to fit the parameters of the regression formulae in the model.
Then, the fitted model with different hyperparameters is used to predict the response variables for the observations in a    validation dataset, with the aim of identifying the    hyperparameters that can   provide a better fit.
Finally,   test datasets are   used to provide an  evaluation of a final model in  the aspects of the coauthor  distribution of researchers,
the evolutionary trend of their number of  coauthors, and the occurrence probability of collaboration events.

%The fitted model is used to predict the responses for the observations in a second dataset called the validation dataset.
%The validation dataset provides an unbiased evaluation of a model fit on the training dataset while tuning the model's

%Validation datasets can be used for the  cumulative advantage of coauthor quantity,
%which does not consider in the training process.

\subsection*{Training}

The parameters of the model are obtained by   two piecewise Poisson models and a log-log model based on a training dataset. Consider a training dataset
consisting of the researchers who produced  publications at the  time interval $[T_0, t_{L-1}]$  and their publications at the time interval $[T_0, t_{L}]$.

% The proportionate $\mathrm{e}^{ {  \beta}_i {  t}_j  }$  changes this quantity.

Firstly,
treating  the index $i$ of $\lambda_{ij}$   as a dummy index,   we assumed $\lambda_{i1}>0$ and
\begin{equation}\lambda_{ij}=\lambda_{i1}   \mathrm{e}^{{ \beta_i} ({t}_j-t_1)  }, \label{eq1}
\end{equation}
   where    $\beta_i $ is the   effect  of time  ${t}_j$.
   Taking logs in   Eq.~(\ref{eq1})
 obtains
\begin{equation}\log \lambda_{ij} = \alpha_i+\beta_i ( t_j-t_1),\label{eq2}
\end{equation} where $\alpha_i=\log \lambda_{i1}$.
 For the majority of researchers who produced $i$ publications  at   $[T_0,t_{j-1}]$,
 their   number of the publications produced  at a following short time interval $(t_{j-1},t_j]$
  follows a Poisson distribution\cite{Xie2019ppp}.
 Therefore,  for each $i\in\{1,2,...,I\}$,
   Eq.~(\ref{eq2}) is the   formula of a one-variable  Poisson   model\cite{Nelder-Wedderburn1972}.

% The linear regression then can be utilized   to calculate $ \alpha_j$ and $\beta_j$.

 %Eq.~(\ref{eq4})   describes  the relationship between the publication creativity   $\lambda_{ij}$ and the covarite $x_j$.

%Now  let us show how to calculate the publication productivity.

Secondly,
treating  the index $j$ of $\lambda_{ij}$   as a dummy index,   we assumed $\lambda_{1j}>0$ and
 \begin{equation}\lambda_{ij}=\lambda_{1j}  i^{\nu_j} , \label{eq3}
\end{equation}
   where
 $\nu_j $ tunes the   effect of  $i$ on $\lambda_{ij}$. The form of Eq.~(\ref{eq3})   is based on   Lotka's law\cite{Xie2019ppp}.
  Taking logs in     Eq.~(\ref{eq3})  obtains
\begin{equation}\log  \lambda_{ij} = \mu_j+\nu_j\log i,\label{eq4}
\end{equation} where $\mu_j=\log \lambda_{1j}$.
For each $j\in\{1,2,...,J\}$,   Eq.~(\ref{eq4}) is the   formula of a log-log model.

% Therefore,  for each $i\in\{1,2,...,I\}$,
%the formula in  Eq.~(\ref{eq2})  restricted at $(t_{j-1},t_j]$

%Therefore, for each  time interval,
%   Eq.~(\ref{eq5})  is a one-variable  log-log   model (see its definition in Appendix A).

Thirdly,
treating  the index $m$ of $\zeta_{mj}$   as a dummy index,   we assumed  $\zeta_{m1}>0$ and
 \begin{equation}\zeta_{mj}=\zeta_{m1}   \mathrm{e}^{{ \varepsilon_m} ({t}_j-t_1)  } , \label{eq5}
\end{equation}
   where
 $\varepsilon_m $ tunes the   effect of time $t_j$ on $\zeta_{mj}$.
  Taking logs in     Eq.~(\ref{eq5})  obtains
\begin{equation}\log \zeta_{mj} = \epsilon_m+\varepsilon_m ( t_j-t_1),\label{eq6}
\end{equation} where $\epsilon_m=\log \zeta_{m1}$.

% For each $m\in\{1,2,...,M\}$,  Eq.~(\ref{eq6})   is the   formula of a Poisson model.

%However,  for the  empirical dataset analyzed here, the    coauthor quantity at the $j$-th time interval attracted by a researcher who produced $m$ publication  at that time interval did not follow a Poisson distribution. Therefore, the value
%of    $\zeta_{mj}$ will be modified.

The fraction of productive researchers and that of the  researchers with many coauthors are small, whereas regression needs enough data.
Therefore, when calculating
$\eta_{ij}$,
we only considered the researchers of the training dataset  whose number of publications at $[T_0, t_{j-1}]$ is no more than a given integer $K$.
When calculating
$\xi_{mj}$,
we only considered the researchers of the training dataset, whose number of publications at $(t_{j-1},t_{j}]$ is no more than a given integer $M$.
  Algorithms  \ref{tab1} and \ref{tab2} are provided to calculate the six parameters in above   formulae based on a training dataset.
% where   $I$  and       $M$  are the largest   publication and annual publication quantities   that we can predict.
%Partition these researchers into $K$ subsets according to their  publication quantity. , where $L$ is an integer larger than $1$
\begin{algorithm}
\caption{Calculating  the matrix   $(\lambda_{ij})_{I\times J}$.}
\label{tab1}
\begin{algorithmic}
%\STATE{{\bf Input}
\REQUIRE ~~\\ %算法的输入参数：Input
the  matrix  $(\eta_{ij})_{K\times L}$. \\
%the  annual coauthor  quantity of researchers: $(k_s(0),k_s(1),...,k_s(t))$; \\
%the  annual publication  quantity of researchers: $(h_s(0),h_s(1),...,h_s(t))$; \\
\ENSURE ~~\\ %算法的输出：Output
\STATE{the matrix  $(\lambda_{ij})_{I\times J}$.}
\FOR{$i$ from $1$ to $K$}
\STATE{replace the $\lambda_{ij}$ in Eq.~(\ref{eq2}) by $\eta_{ij}$  for $j= 1,...,L$; }
\STATE{calculate $ \alpha_i$ and $\beta_i$ by   the linear regression;}
\STATE{let  $\lambda_{ij}=  \mathrm{e}^{\alpha_i+{ \beta_i} ({t}_j-t_1)}  $ for $j=1,...,J$;   }
\ENDFOR
\FOR{$j$ from $1$ to $L$}
\STATE{replace the $\lambda_{ij}$ in Eq.~(\ref{eq4}) by $\eta_{ij}$  for $i=1,...,K$; }
\STATE{calculate $ \mu_j$ and $\nu_j$ by the linear regression;}
\STATE{let  $\lambda_{ij}=  \mathrm{e}^{\mu_j}  i^{\nu_j} $  for $i=K+1,...,I$;   }
\ENDFOR
\FOR{$i$ from $K+1$ to $I$}
\STATE{replace the $\lambda_{ij}$ in Eq.~(\ref{eq2}) by  $  \mathrm{e}^{\mu_j}  i^{\nu_j}   $  for $j=1,...,L$; }
\STATE{calculate $ \alpha_i$ and $\beta_i$  by the linear regression;}
\ENDFOR
\FOR{$j$ from $L+1$ to $J$}
\STATE{replace $\lambda_{ij}$ in Eq.~(\ref{eq4}) by    $   \mathrm{e}^{\alpha_i+{ \beta_i} ({t}_j-t_1)}     $ for $i=1,...,K$; }
\STATE{calculate $ \mu_j$ and $\nu_j$  by the linear regression;}
\ENDFOR
\STATE{let  $\lambda_{ij}=(\mathrm{e}^{\alpha_i+{ \beta_i} ({t}_j-t_1)}  + \mathrm{e}^{\mu_j}  i^{\nu_j}  )/2$ for  $i=K+1,...,I$ and
$j=L+1,...,J$.}
% \RETURN  {the matrix  $(\lambda)_{I\times J}$.}
\end{algorithmic}
 \end{algorithm}

\begin{algorithm}
\caption{Calculating  the matrix   $(\zeta_{mj})_{M\times J}$.}
\label{tab2}
\begin{algorithmic}
%\STATE{{\bf Input}
\REQUIRE ~~\\ %算法的输入参数：Input
the  matrix $(\xi_{mj})_{M\times L}$. \\
\ENSURE ~~\\ %算法的输出：Output
\STATE{the matrix  $(\zeta_{mj})_{M\times J}$.}
\FOR{$m$ from $1$ to $M$}
\STATE{replace the $\zeta_{mj}$ in Eq.~(\ref{eq7}) by $\xi_{mj}$   for $j=1,...,L$; }
\STATE{calculate $ \tau_m$ and $\upsilon_m$ by   the linear regression;}
\STATE{let  $\zeta_{mj}=  \mathrm{e}^{\upsilon_m+{ \tau_m} ({t}_j-t_1)}  $ for $j=1,...,J$.   }
\ENDFOR
% \RETURN  {the matrix  $(\lambda)_{I\times J}$.}
\end{algorithmic}
 \end{algorithm}

  Note that the training dataset would not contain enough productive researchers. It would cause that the parameter $K$  is much smaller than the largest  number publications $I$ that the model can
 predict.
In this case, the model will give   bad prediction results
to productive researchers.
%Therefore, a large training dataset is needed.

\subsection*{Validating}

% only use the information of annual average coauthor quantity., but a generalized one

 The
hyperparameters of the model are obtained
 based on a validation dataset.
 Eq.~(\ref{eq6}) is the   formula of a   Poisson model.
However,   the  number of  coauthors at the $j$-th time interval of a researcher in the considered dataset (who produced $m$ publication  at that time interval) did not follow a Poisson distribution (see Section 4).
Therefore, the value of  $(\zeta_{mj})_{M\times J}$  calculated by a Poisson model should be modified.
Consider a validation dataset consisting of the researchers who produced  publications at the  time interval  $[t_U,t_{U+1})$.  Consider  their annual number   of  publications   at  $[T_0,  t_{V}]$, where $t_U<t_V$.

Consider a researcher $s$ in the validation dataset. Consider
     the series of his or her number of   coauthors    and that of publications   $(k_s(t_U),...,k_s(t_{V}))$ and  $(h_s(t_U),...,h_s(t_{V}))$, where
 $k_s(t_l)$ and $h_s(t_l)$ are the number of his  or her  coauthors and the number  of his  or her publications at $[T_{0},t_{l}]$ ($t_U\leq t_l \leq t_V$).
The formula of his or her $\zeta_{mj}$ is modified   as
 \begin{equation}
(\tilde{\zeta}_{mj})_s=
\begin{cases}
\upsilon   {\zeta}_{mj}, & \text{if $ k_s(t_{j-1})=0$}, \\
\upsilon k_s(t_{j-1})^\tau {\zeta}_{mj},  & \text{if $ k_s(t_{j-1})>0$},
\end{cases}
\label{eq7}
\end{equation}
 where $\upsilon$ and   $\tau>0$.
 The formulae in Eq.~(\ref{eq7}) express the  cumulative advantage of attracting new coauthors on researchers' historical number of coauthors.

In the training process, $\upsilon=1$ and $\tau=0$. Choosing different values of $\upsilon$ and $\tau$ will obtain different prediction results;
 thus they can be regarded as   hyperparameters.
The explicit formulae of $\upsilon$ and $\tau$ cannot be obtained; thus
Algorithm~\ref{tab3} is proposed   to calculate them for a good fitting to the validation dataset, which is a genetic algorithm.

%Partition these researchers into $N$ subsets according to their history coauthor quantity.

%$(k_s(t_U),...,k_s(t_V))$ and $(h_s(t_U),...,h_s(t_V))$  $s$

\begin{algorithm}
\caption{Calculating  the hyperparameters in Eq.~(\ref{eq7}).}
\label{tab3}
\begin{algorithmic}
%\STATE{{\bf Input}
\REQUIRE ~~\\ %算法的输入参数：Input
the series $(k_s(t_U),...,k_s(t_{V}))$ and  $(h_s(t_U),...,h_s(t_{V}))$
of    any  researcher $s$ in the validation dataset; \\
the    matrix  $(\zeta_{mj})_{M\times J}$;\\
the parameters   $n_k$ ($k=0,..,3$) and  intervals $L_l$ ($l=0,...,2$).\\ %}
\ENSURE {the first chromosome.}\\ %算法的输出：Output

\STATE{initialize   a randomly generated population of $n_0$ chromosomes: $(\tau,  \upsilon)\in L_0\times L_1$;}
\REPEAT
   \STATE{//create   chromosomes: crossover}
\REPEAT
\STATE{  select a pair of parent chromosomes   ($\tau_1$, $\upsilon_1$) and  ($\tau_2$, $\upsilon_2$) randomly;}\\
\STATE{  generate a random number $r\in [0,1]$;}\\
\STATE{  generate a   chromosome ($r\tau_1+(1-r)\tau_2$, $r\upsilon_1+(1-r)\upsilon_2$);}\\
 \UNTIL { $n_1$ times }
   \STATE{//create chromosomes: mutation}
   \REPEAT
\STATE{  select a   chromosome    ($\tau$, $\upsilon$) randomly;}\\
\STATE{  generate two random numbers $r_1,r_2\in L_2$;}\\
\STATE{  generate a   chromosome ($\tau+r_1$, $\upsilon+r_2$);}\\
 \UNTIL { $n_2$ times }
\STATE{let  $\Delta h_s(t_l)=h_s(t_l)-h_s(t_{l-1})$ and $\Delta k_s(t_l)=k_s(t_l)-k_s(t_{l-1})$;}
   \STATE{calculate  fitness:  $f(\tau,\upsilon)=\sum_{s,l}\big| ( {\tilde{\zeta}}_{\Delta h_s(t_l)t_l})_s-   \Delta k_s(t_l)  \big|    $,;}

  \STATE{select the first  $n_0$   chromosomes $\in L_0\times L_1$    according to the ascending  order of fitness.}
 \UNTIL { $n_3$ times }
\end{algorithmic}
 \end{algorithm}

\subsection*{Testing}

 Consider a test dataset consisting of the researchers who produced publications at the time interval
$[t_X,t_{X+1})$, the historical  number of their publications  and
 the number of their coauthors
at the time interval $[T_{0},t_{X}]$,   the  annual number of their publications  and
the number of their new coauthors
at the time interval $[t_{Y},t_{Z}]$, where  $t_X<t_Y<t_Z\leq t_J$.
Due to the data-size requirement
of using regression, we  only predicted  the number of coauthors  for
the researchers with    annual number of publications    no more than  $M$
and historical  number of publications    no more than  a given integer $I_1$.

 %Some historical information of those researchers are used in training and validation.
%The test dataset is used in the two processes.

Note that the annual number of new  coauthors  depends on the annual  number of publications   and   the number of new coauthors  in each publication, namely two random variables.
This is   modelled  by
 Algorithm~\ref{tab4}.
 Due to its regression nature,  this algorithm cannot predict the exact number of publications  for an individual, but   can be suitable  for a group of researchers.

% and he new coauthors of each publication also follows such a distribution.
%Therefore, the annual newly coming  coauthors is
%a composite of Poisson process.

\begin{algorithm}
\caption{Predicting the number of publications and that of coauthors for researchers.}
\label{tab4}
\begin{algorithmic}
%\STATE{{\bf Input}
\REQUIRE ~~\\ %算法的输入参数：Input
the $h_s(t_X)$ and $k_s(t_X)$ of any   researcher $s$ in a test dataset;
 \\the matrixes $(\lambda_{ij})_{I\times J}$ and $(\zeta_{mj})_{M\times J}$; %the predicted publication creativity and cooperation activity
\\
the hyperparameters  $\upsilon$ and $\tau$. %}
\ENSURE ~~\\ %算法的输出：Output
the   $h_s (t_Z)$ and   $k_s (t_Z)$ of any researcher $s$.
\FOR{   each researcher $s$}
\STATE{initialize $h=h_s(t_X)$ and $k=k_s(t_X)$;}
\FOR{$l$ from $X+1$ to $Z$}
\STATE{sample an integer  $r$ from   Pois$(\lambda_{hl})$; }
\STATE{sample an integer  $u$ from Pois$(  (\tilde{\zeta}_{rl})_s)$; }
\STATE{let  $h= h    + r$ and  $k= k    + u$;}
\ENDFOR
\STATE{let $h_s (t_Z)=h$ and $k_s (t_Z)=k$.}
\ENDFOR
%\RETURN  { $h_s (t_Z)$ and  $k_s (t_Z)$.}
\end{algorithmic}
 \end{algorithm}

\section*{Results}

\subsection*{Empirical  data}

 The training, validation, and test datasets of our study are  extracted from the dblp dataset (Table~\ref{tab5}), in which
  the publications with more than 80 authors have been  filtered.
 Sets 1 and 2 are used to extract the historical number of publications for the researchers in  Sets 3 and 4.
Set   5 is used as a training dataset,
Set 6 is used as a validation  dataset, and
Sets 7 and 8 are used to test  the prediction results for the researchers in  Sets 3 and 4.
These datasets   consist of 220,344 publications  in 1,586 journals and conference proceedings that were produced by  328,690 researchers  from 1951 to 2018.
Due to the size of the analyzed datasets, the proposed  model is applicable at least to   researchers in computer science.

%Here, we considered a cleared  dataset  spanning a long time interval.

%extracted parts of the data in certain time intervals as   which are the same those in Reference~\cite{xie2019ppp}.

\begin{table}[!ht] \centering \caption{{\bf Considered subsets  of the dblp   dataset.} }
\begin{tabular}{l ccccccccccc} \hline
Dataset&  $a$   & $b$ &  $c$  & $d$ & $e$ & $f$    \\ \hline
Set 1 &1951--1994 &   180,45& 18,398& 319& 1.558 &1.528 \\
 Set  2   &1951--2000 &  38,149& 35,643& 542 &1.571& 1.681\\
   Set   3   &1994 &2,903  &  1,922& 146 &  1.137  &  1.718 \\
 Set  4  &2000 & 5,741   &   3,600& 257  & 1.184 & 1.888 \\
% Set  5 & 1994--2009 & 88,853&64,558& 940& 1.545 &    2.126\\
 Set  5 & 1985--2009 & 97,321&75,338& 964& 1.591 &     2.055 \\
Set  6 & 2000--2009 & 73,642 &48,991& 874&1.480&    2.224\\
   Set   7 &  1995--2018 &316,212& 201,946 &1,538  & 1.754& 2.746\\
 Set   8 &  2001--2018 &  301,741& 184,701 & 1,495 &1.733& 2.831 \\
 \hline
 \end{tabular}
  \begin{flushleft} The index  $a$:    the time interval of data,  $b$: the number  of researchers, $c$: the number of publications,  $d$: the  number of journals,  $e$: the average number of publications of researchers, $f$: the average number of authors of publications.
\end{flushleft}
\label{tab5}
\end{table}

%\begin{table}[!ht] \centering \caption{{\bf Training, validation,   and test  datasets.} }
%\begin{tabular}{l lcccccccccc} \hline
%Dataset&  $a$ &   $b$ &  $c$  & $d$ & $e$ & $f$    \\ \hline
%Set 1 &1951 ($T_0$)--1985 ($T_1$, $t_0$)   & 6,285&  5,099& 132& 1.592& 1.293 \\
%Set  2   &1951 ($T_0$)--2000 ($t_U$, $t_X$)   & 35643& 38149& 542 &1.571 & 1.681 \\
%  Set  3 & 1985 ($T_1$, $t_0$)--2009 ($t_L$)   &75338&97321& 964& 1.591  & 2.055  \\
%  Set  4  &2000 ($t_U$, $t_X$)  & 5,741   &   3,600& 257  & 1.184 & 1.888 \\
% Set   5 &  2001 ($t_{U}$)--2009 ($t_V$) &  45,391& 70,115& 860 &1.457 & 2.251  \\
% Set   6 &  2010 ($t_{Y}$)--2018 ($T_2$, $t_Z$ )   & 139,310& 251,999 & 1,414 & 1.669& 3.020\\
%\hline
% \end{tabular}
%  \begin{flushleft} The index  $a$:    the time interval of data,  $b$: the number  of researchers, $c$: the number of publications,  $d$: the  number of journals,  $e$: the average number of researchers' publications, $f$: the average number of publications' authors.
% \end{flushleft}
%\label{tab5}
%\end{table}

The parameters of the training dataset Set 5 are $I=180$,  $J=33$,  $K=42$, $L=24$, $M=12$, $T_0=1951$, $T_1=t_0=1985$,    $t_L=2009$,  and $t_{J}=T_2=2018$.
In detail, it  consists of the researchers who have publications at   $[T_0,t_{L-1}]$ and their annual number of publications   at $[T_0,t_L]$.   Due to the low bound of data size of using regression,
 we   only consider
  the researchers with   no more than  $K$ publications
and those with    no more than  $M$   publications at   $(t_{0},t_{L}]$.
  Algorithms~\ref{tab1} and \ref{tab2}   are provided to calculate   $(\lambda_{ij})_{I\times J}$ and   $(\zeta_{mj})_{M\times J}$
based on
the  matrixes
$(\eta_{ij})_{K\times L}$   and
    $(\xi_{mj})_{M\times L} $ that are calculated on the basis of    the training dataset.

% for  $i=1,...,I$ and $j=1,...,J$
%For example,
% $\eta_{11}$  is average publication quantity  at the year $1986$
%  of researchers with one publication at  the years  from $1951$ to  $1985$.
% And  $\xi_{11}$   is average coauthor quantity  at the year $1986$
%  of researchers with one publication at  the year $1986$.

The parameters of  validation dataset (Set 6)  and the  test dataset  (Set 4)  used here are    $t_U=t_X=2000$, $t_V=2009$, $t_Y=2010$, and
  $t_Z=2018$. In detail,
the validation dataset  consists of the researchers  who have publications at the time interval  $(t_{U-1},t_{U}]$    and their annual number of publications at  $[T_0,t_V]$.
It is used to calculate the hyperparameters $\upsilon$ and $\tau$ by Algorithm~\ref{tab3} with the parameters:
    $n_0=400$, $n_1=0.6 n_0$,  $n_2=0.3 n_0$,   $n_3=500$,
$L_0=[0.6,1.0]$, $L_1=(0.0,0.4]$, and
$L_2=[-0.01,0.01]$.
  The results are $\upsilon=0.603$ and $\tau=0.321$.

%0.193300000000000	0.710600000000000
%0.219900000000000	0.733700000000000
%0.198600000000000	0.806000000000000
%0.168700000000000	0.890800000000000
%0.183000000000000	0.694000000000000

  The test dataset  consists of the researchers  who produced publications at  $[t_X,t_{X+1})$    and the  annual number of their publications and the number of their   new coauthors at  $[t_Y,t_Z]$.
 We   predicted the two variables     only for  99.96\% of the  researchers  in Set 4,   who have no more than  $I_1=40$ publications at   $[T_0,t_X]$ and no more than $M=12$ annual publications at $[t_X,t_Z]$.

\subsection*{The reasonability  of the model assumptions}
\begin{figure*}[h]
\centering
% Use the relevant command to insert your figure file.
% For example, with the graphicx package use
\includegraphics[height=2.3   in,width=4.6     in,angle=0]{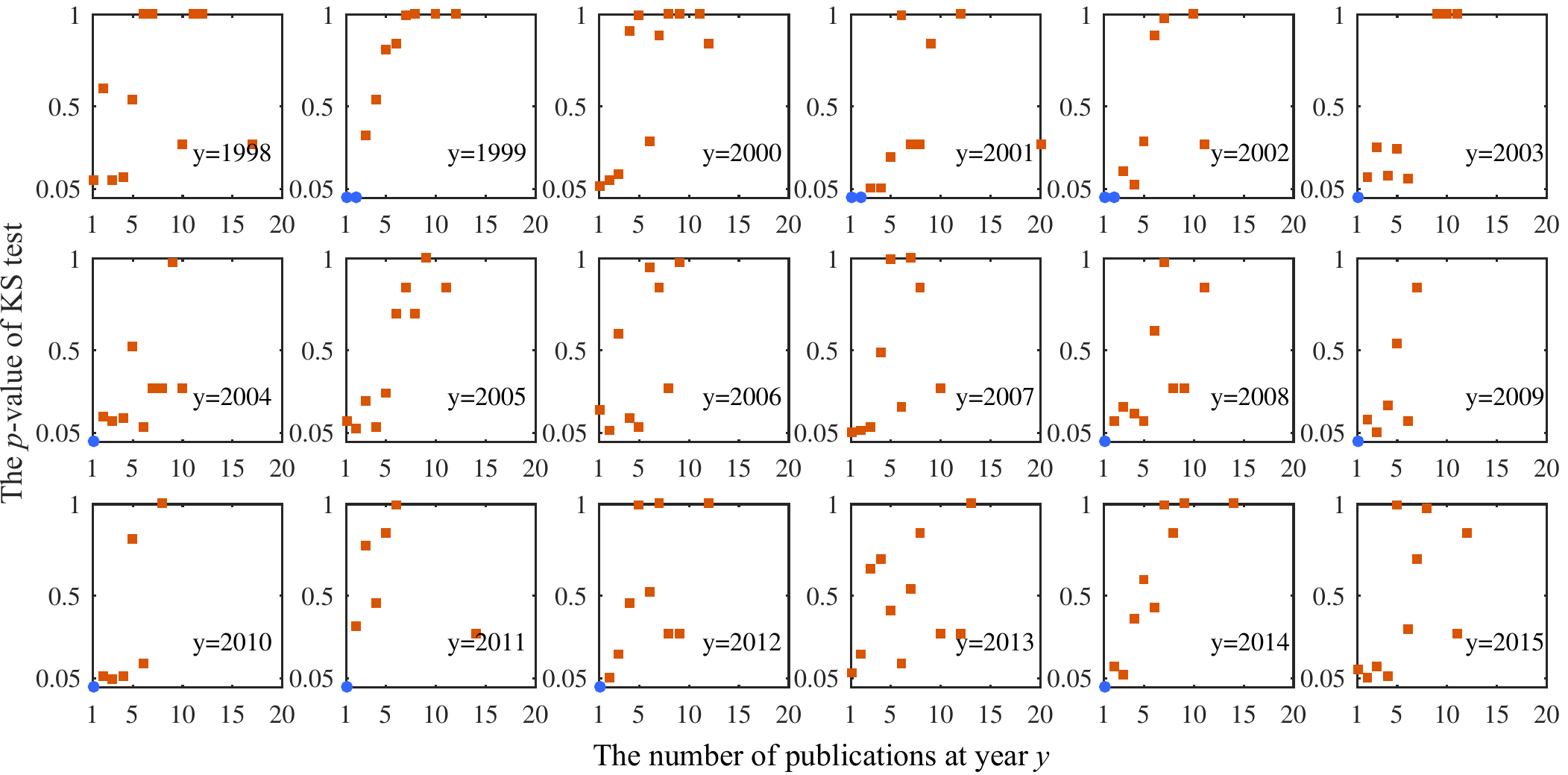}
% figure caption is below the figure
\caption{    {\bf  The $p$-value of the KS test with the hypothesis that a random variable follows a   Poisson   distribution.}
 The panels show
 the  results  of the KS test on the number of coauthors of   a researcher   with   $m$ publications   in   year   $y$,   where $m=1,...,20$.
 If the $p$-value$\leq 0.05$,   the test rejects the hypothesis   (blue circles); otherwise, it cannot reject the hypothesis (red squares).
}
 \label{fig1}
\end{figure*}

\begin{figure*}[h]
\centering
% Use the relevant command to insert your figure file.
% For example, with the graphicx package use
\includegraphics[height=2.5   in,width=4.6     in,angle=0]{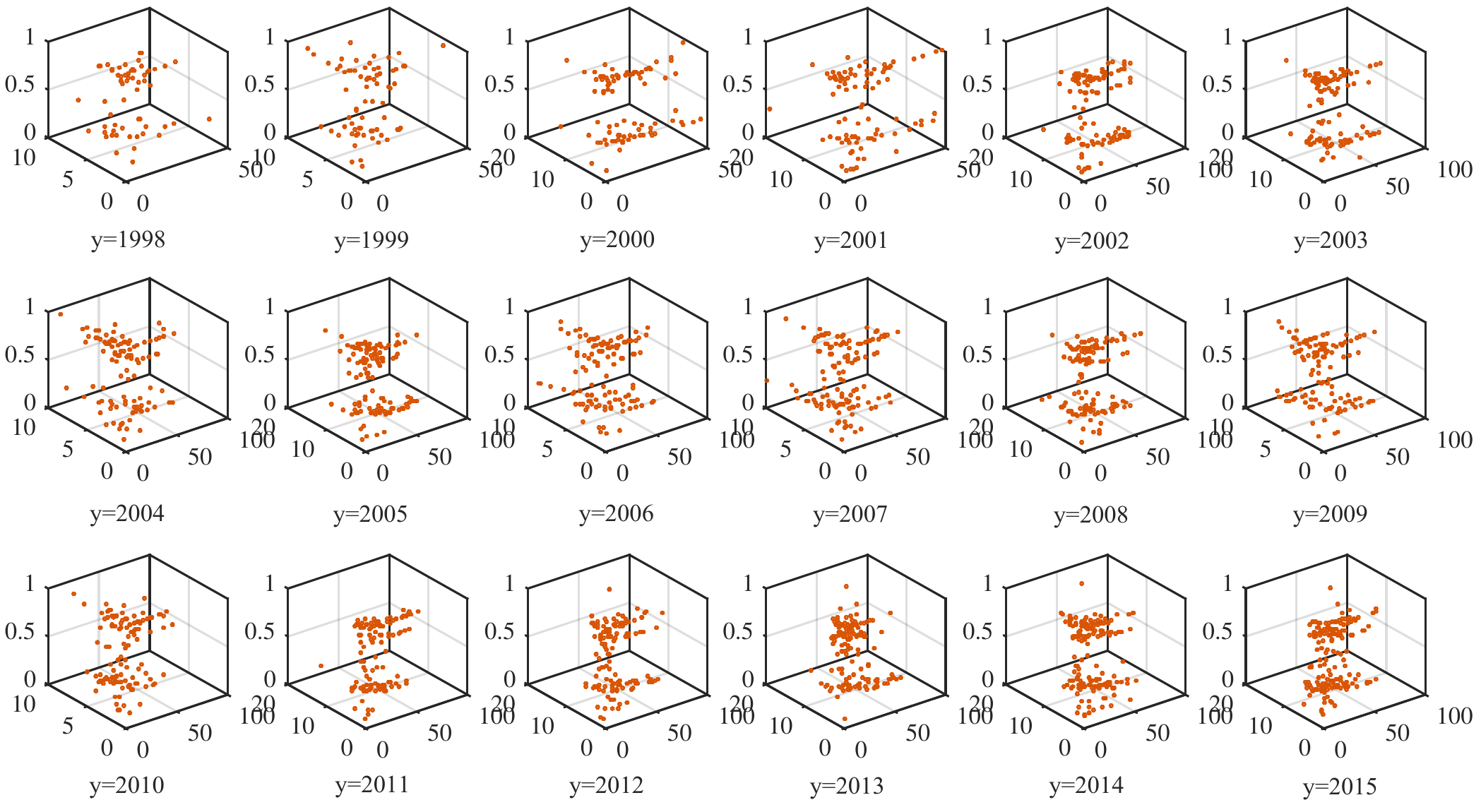}
% figure caption is below the figure
\caption{    {\bf Eliminating the diversities in  the historical  number of  coauthors and the annual number of publications   induces    Poisson   distributions.  }
 Consider the
researchers with $m$ publications (upper right direction)  in year $y$   and with $l$ coauthors before $y$ (upper left direction).
If the $p$-value$\leq 0.05$  (vertical direction),   the KS test rejects the hypothesis  that the number of  coauthors of that   researcher  follows a   Poisson   distribution (blue squares); otherwise, it cannot
reject the hypothesis (red circles).}
  \label{fig2}
\end{figure*}

First, we showed the reasonability of modifying  $(\zeta_{mj})_{M\times J}$.
We partitioned   the researchers of  the training dataset into subsets according to their    number of  publications in a given year.
 The   Kolmogorov-Smirnov (KS) test  rejected the null hypothesis that the coauthor distributions   of some large    subsets (with 1 or 2 annual  publications)
   are    Poisson distributions (Fig.~\ref{fig1}).
Diminishing the diversity in researchers' historical number of coauthors   reveals   the Poisson character  of the coauthor distributions.
The  KS  test cannot reject the null hypothesis that the coauthor distributions   of    researchers
with the same   annual number of  publications  and historical number of coauthors
are Poisson distributions (Fig.~\ref{fig2}).
Therefore, it is  necessary    to modify  $(\zeta_{mj})_{M\times J}$ to express  the cumulative advantage of  the historical number of coauthors, which gives the reasonability
of the formulae in Eq.~(\ref{eq7}).

%cannot reject
%the quantitative distribution of  coauthors at the year $y$ produced by
%  the researchers have $m$ publications at the year   $y-1$  is a   Poisson   distribution.  Panels show
% the $p$-value of the KS  test .

%  \begin{figure*}[h]
%\centering
%% Use the relevant command to insert your figure file.
%% For example, with the graphicx package use
%%\includegraphics[height=0.9    in,width=2.1     in,angle=0]{Figure-12.pdf}
%% figure caption is below the figure
%\caption{    {\bf   Poisson feature holds for the majority of   researchers.  }
%Red circles show
% the $p$-value of the KS test  in Fig~\ref{fig3} only on the researchers with no more than six publications per year.
%Index $q$ is proportion of the researchers whose
%   subset      can pass  the KS test. }
% \label{fig12}
%\end{figure*}

%For each year $y$, we counted  the number of those researchers whose subset

%   count over the total number of researchers who produced publication at the year $y$.

\begin{figure*}[h]
\centering
% Use the relevant command to insert your figure file.
% For example, with the graphicx package use
\includegraphics[height=1.55   in,width=4.6    in,angle=0]{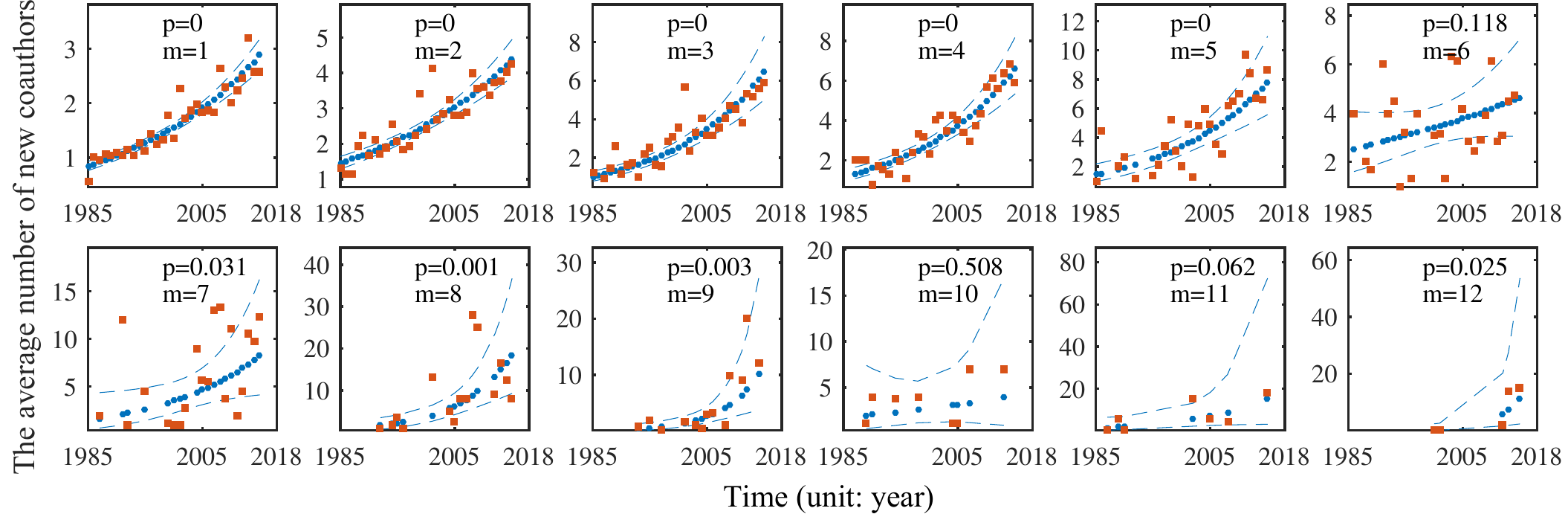}
% figure caption is below the figure
\caption{   {\bf  The relationship between  $\xi_{mj}$   and    $t_j$ given $m$.} Consider the researchers with      $m$ publications  at $(t_{j-1},t_j]$.
 The panels show  the  average number of   new coauthors of these researchers   at $(t_{j-1},t_j]$   ($\xi_{mj}$,  red squares),
the predicted results by the    Poisson regression  ($\zeta_{mj}$, blue dots),  and   the confidence intervals of the regression (dashed lines).
When  $p<0.05$, the $\chi^2$  test    rejects the null hypothesis that the regression coefficient of   time is equal to zero.  }
 \label{fig3}
\end{figure*}

%: their member's  annual publication quantity follows a Poisson distribution
Secondly, we showed the significance  of the  regression results on the
training dataset. The $\chi^2$ test  indicates that $\eta_{ij}$    significantly correlates to  $i$  given  $j$,  and  to  $j$   given  ${i}$\cite{Xie2019ppp}.
 The $\chi^2$ test  indicates that  $ \xi_{mj}$ significantly correlates to  $t_j$ given  $m$ from $1$ to $9$ except $6$ (see the $p$-value   in Fig.~\ref{fig3}).
The researchers with that $m$ account  for  $99.68\%$ of the researchers in the training dataset.
  These significant correlations guarantee the effectiveness of  utilizing   regression methods   to   calculate   $(\lambda_{ij})_{I\times J}$ and  $(\zeta_{mj})_{M\times J}$.

\subsection*{Evolutionary  trend of the number of   coauthors}
 %We analyzed the correlation between the empirical data and the predicted data. Note that  the index $j$ is treated as a dummy index.

\begin{figure*}[h]
\centering
% Use the relevant command to insert your figure file.
% For example, with the graphicx package use
\includegraphics[height=2.2   in,width=4.6    in,angle=0]{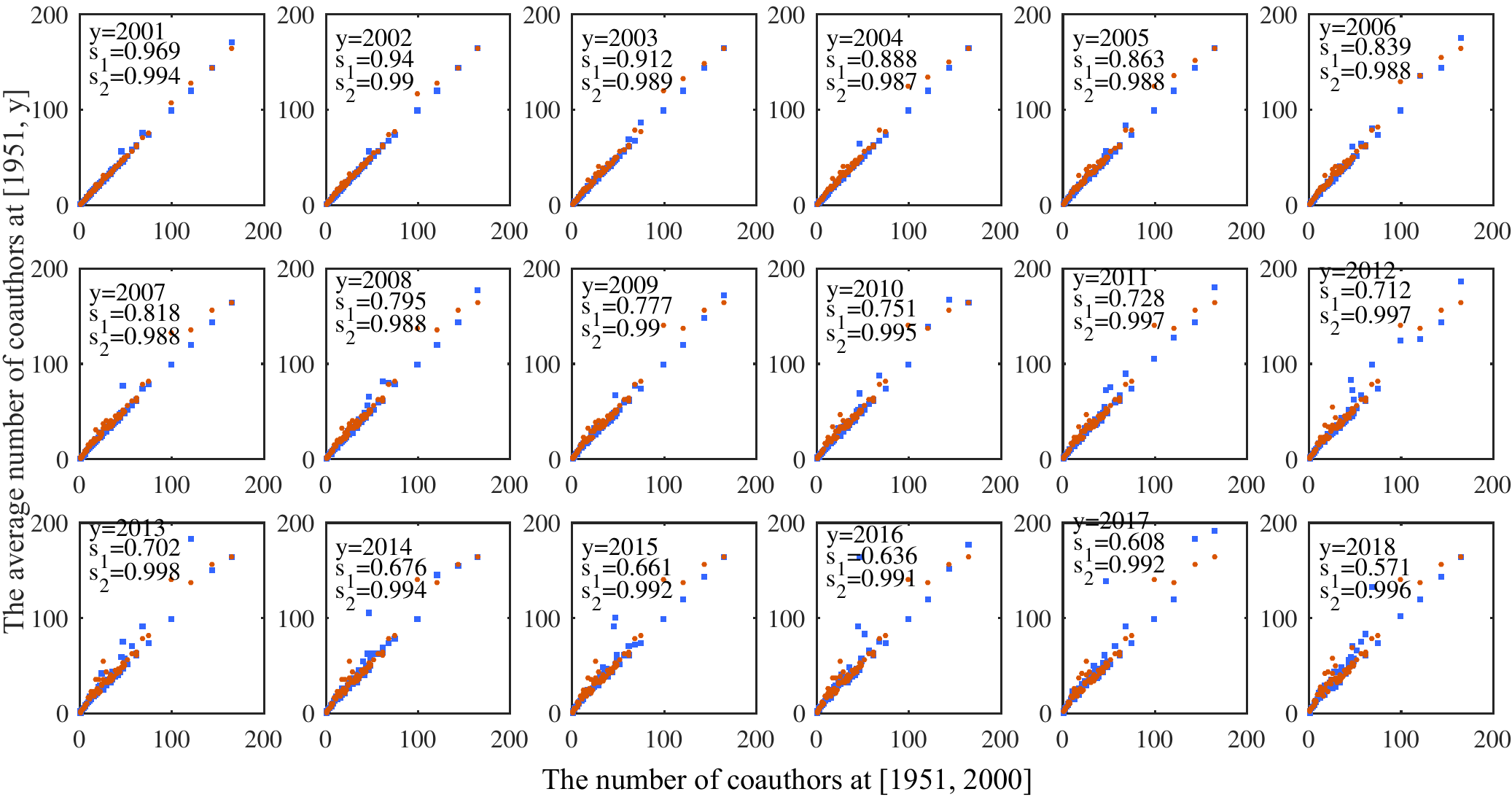}
% figure caption is below the figure
\caption{   {\bf  Fittings on the evolution of   the number of researchers' coauthors.} Consider the tested   researchers   who have  $k$ coauthors    at   $[1951,2000]$, where $k=1,...,200$.
The panels show
the average number of coauthors   of   these researchers at    $[1951,y]$ ($n(k,y)$,  red dots) and the predicted number    ($m(k,y)$, blue lines).
  Index $s_1$ is the  Pearson  correlation coefficient
     calculated
 based on  the  list of researchers'  number of coauthors and   their predicted number.
   Index $s_2$ is this  coefficient based on the sorted
    lists.}
 \label{fig4}
\end{figure*}

Consider    the tested researchers   who have   $k$  coauthors at the time interval $[T_0,t_X]$.
  Let $n(k,t_j)$ be the average number of these researchers'  new coauthors  arriving   at the time interval $(t_{j-1},t_{j}]$, and $m(k,t_j)$ be that predicted by the model.
  Fig.~\ref{fig4} shows the   trends of $n(k,t_j)$ and $m(k,t_j)$  on $k$ at each year $t_j$  from $2001$ to $2018$.

 The correlation of     the trends  is measured by the  Pearson correlation coefficient\cite{Hollander} on individual level ($s_1$: calculated
 based on  the  list of researchers'   number of  coauthors and that of their predicted one) and that on
 group level ($s_2$: sort the lists, and then calculate the coefficient).  The value of $s_1$ decreases over time, whereas that of $s_2$ keeps high.
  It indicates that  the model is unapplicable to    the long-time prediction for individuals, but can be applicable for a group of researchers.

%the prediction precision for individuals decreases over time, which means

%Note that the   coefficient
%indicates the strength of a linear
%relationship between two variables $x$ and $y$ (which are
%the ranks of analyzed variables), unless   the conditional expected value of $y$
%given $x$ is   linear or approximate linear
%in $x$,  and verse vice. The effectiveness of the correlation analysis here is guaranteed by
%the visual examinations   shown in Fig.~\ref{fig6}.

%\begin{figure}\centering
%% Use the relevant command to insert your figure file.
%% For example, with the graphicx package use
%%\includegraphics[height=1.1  in,width=4.6    in,angle=0]{Figure_2.pdf}
%% figure caption is below the figure
% \caption{{ \bf  The  increasing trends of  coauthor quantity   and publication quantity. }
% The red dots show average coauthor quantity (abscissa) and average publication quantity (ordinate)
%The   correlation can be measured by
%  the  Pearson's linear correlation coefficient $r$, where  $r>0$ means   positive   correlation.
% }
% \label{fig4}      % Give a unique label
%\end{figure}

%(denoted  by $\mathbb{E}(y|x)$)

% (where those publications are produced at the time interval $[T_0, t_j]$)
% for any possible $i$ and $j$
\subsection*{Coauthor distributions}

\begin{figure*}[h]
\centering
% Use the relevant command to insert your figure file.
% For example, with the graphicx package use
\includegraphics[height=2.2   in,width=4.6     in,angle=0]{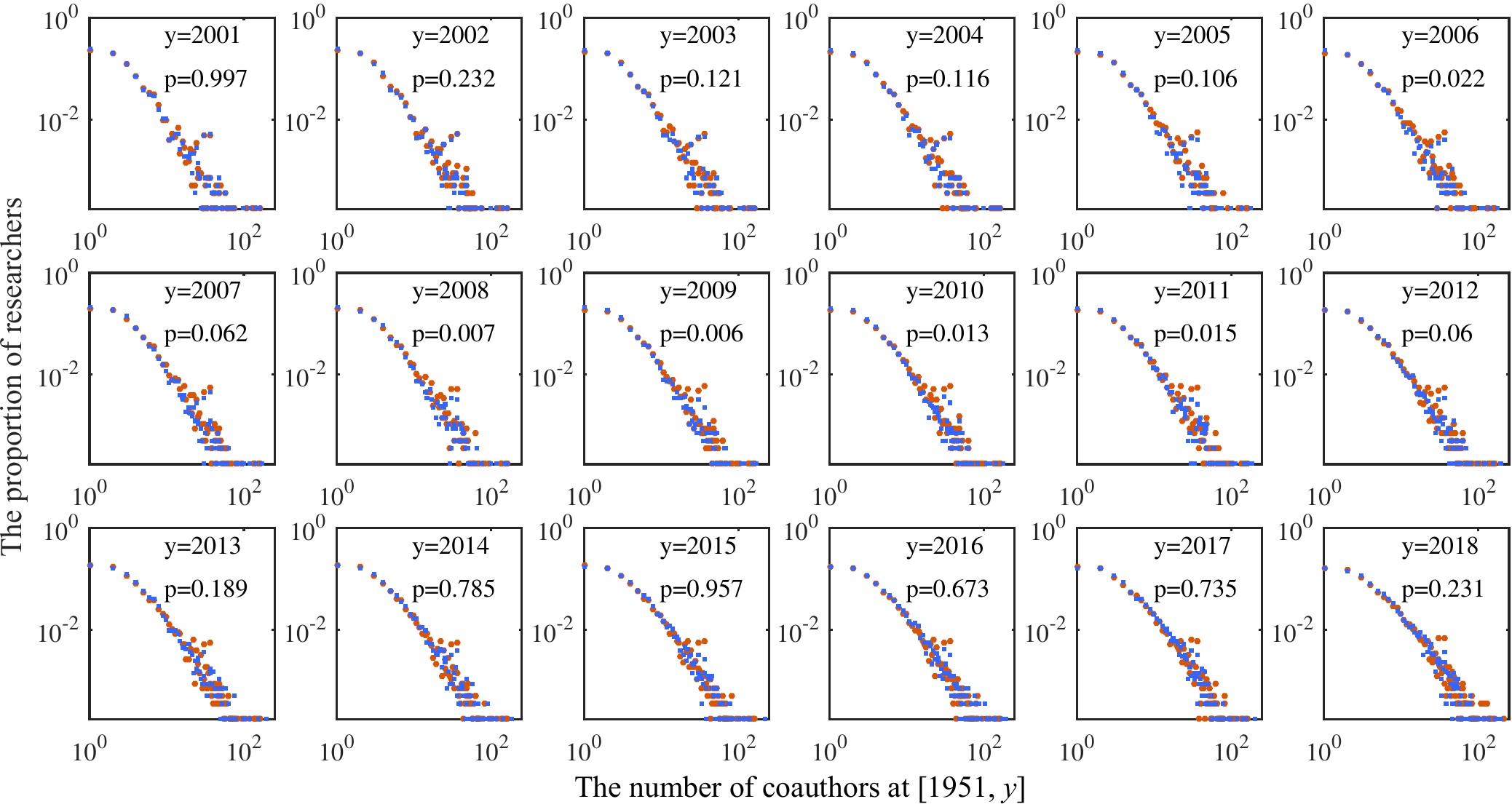}
% figure caption is below the figure
\caption{   {\bf  Fittings on  coauthor   distributions.}
The panels show the coauthor    distribution  of
 the tested   researchers
 at time interval $[1951 ,y]$
    (red circles)  and the  predicted distribution
  (blue squares). When $p>0.05$,   the KS test cannot reject the null hypothesis that the two  distributions are  the same.  }
 \label{fig5}
\end{figure*}

 We  compared the  coauthor distribution of   the tested researchers at $[T_0,y]$   with the    predicted  distribution, where $y= 2001,..., 2018$.
Fig.~\ref{fig5} shows that
a fat tail emerges in the evolution of
the  ground-truth distribution  and in that of the predicted distribution.
This shows that    our model can  capture the
fat-tail phenomenon.
The KS test   rejects that some of the  compared  distributions are the same  (see the $p$-value   in Fig.~\ref{fig5}), although there is a coincidence in their heads. This indicates that the prediction precision   for   researchers
with many coauthors    needs to be improved.

\subsection*{Collaboration events  }
%The correlation between coauthor quantity and publication quantity
% expected number  of  the researcher's publications    produced at the next time interval.

%, where  the  probability larger than 0.5 is
%regarded  as  collaboration, and   smaller as not
The
above two experiments focus on the  prediction precision of our model  over a long time interval.
The following experiment is designed  to test   the   precision over a short time interval, namely, the next year.
The    model  can   provide the
 probability of  the researcher $s$ having new coauthors in the next time interval $(t_{l-1},t_l]$:
\begin{equation}p_s(t_l)=  1- \mathrm{ e}^{-\lambda_{h_s(t_{l-1})t_l}} -  \sum^M_{x=1}\frac{x^{\lambda_{h_s(t_{l-1})t_l}}}{x!}\mathrm{e}^{-\lambda_{h_s(t_{l-1})t_l}}
\mathrm{ e}^{-(\tilde{\zeta}_{xt_l})_s}
.\label{eq8}
\end{equation}

 The area under the curve (AUC) of the receiver operating characteristic is used to measure the  prediction precision.
 Count the times that a   researcher  did (did not)   collaborate with   new coauthors in the next time interval,
 the probability is larger (smaller)  than 0.5.
Denote the counts by  $m_1$ and $m_2$ respectively.
Count  the times that the probability is 0.5, and denote the count by $m_3$. Denote the number of   tested  researchers by $m$.
 Then,
\begin{equation}\mathrm{AUC}= \frac{m_1+m_2+0.5m_3}{m}.\label{eq9}
\end{equation}

\begin{figure*}[h]
\centering
% Use the relevant command to insert your figure file.
% For example, with the graphicx package use
\includegraphics[height=2.3   in,width=4.6     in,angle=0]{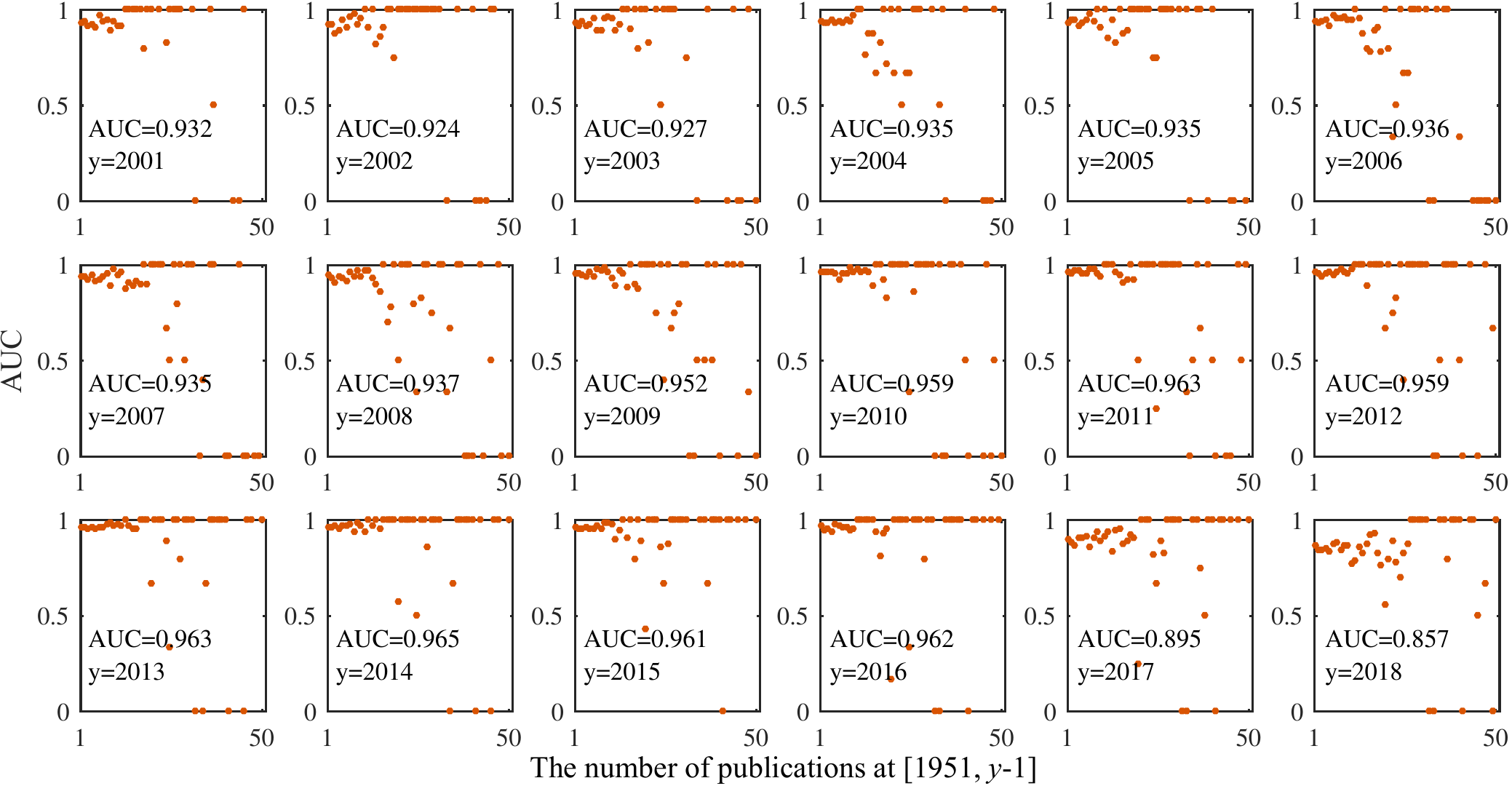}
% figure caption is below the figure
 \caption{        {\bf  The precision of  predicting  collaboration  events.}
 The red dots show
   the   AUC    of  predicting the collaboration events  at     year $y$ for  the tested researchers   who produced  $i$ publications at $[1951,y-1]$, where $i=1,...,50$.  The index AUC is calculated based on all of the tested
researchers.           }
    \label{fig6}
\end{figure*}

Fig.~\ref{fig6}   shows that      the   AUC value is   high  for researchers with a  small historical number of publications. This indicates that the model can provide a satisfactory  prediction for the collaboration events of researchers
with low productivity.  It also indicates that there is no regularity
of collaborations that can be revealed by our model for   highly productive  researchers, which indicates a  direction for improving of the model. Due to the
vast number of low productivity researchers,
the  AUC value is high  for  all of the tested
researchers.

\section*{Discussion and conclusions}
  A learning model is proposed  to predict  the number of coauthors for researchers. Its   practicability is tested on the dblp dataset, and its effectiveness is exhibited by the satisfactory   fittings on the evolutionary trend of the number of coauthors for researchers,  the    distribution of this variable, and the occurrence probability of collaboration events. Note that our model cannot provide an exact prediction for an  individual.
However, due to its nature of regression,
it can still be of use in its ability to provide a satisfactory prediction for a group of randomly selected researchers on average.

The parameters of our model  are learned from a training   dataset, the  methods of which can be generalized to determine the parameters for   models of coauthorship networks or other   network models.
 The hyperparameters of our model  are used   to modify the
  intermediate results   given by regression.
The formulae of modification
 express  the   cumulative advantage of attracting coauthors on  the historical number of coauthors, which enables  our model to directly  predict the number of coauthors generated by an inhomogeneous Poisson process.

%They are calculated here
% by  a  genetic algorithm   based on a validation dataset.

%The coauthor quantity is   dependent on the annual publication quantity and the new coauthor quantities of publications. Using the Poisson model needs to partition researchers into subsets diminishing the diversities in   historical coauthor quantity and     annual publication quantity. However, some of these subsets are too small to use regression.

% Due to the randomness of  individuals' behaviors,
%predicting the coauthor quantity of  an individual would not be done mainly by regression as this study did for a group of researchers.
% The randomness is displayed    by the relatively small   autocorrelation coefficients of the time series on a researcher's cumulative coauthor quantity (see Appendix B).

The model   provides a platform to utilize multiple factors by substituting them in  the right side of Eqs.~(\ref{eq2},\ref{eq4},\ref{eq6}).
 A limitation of the results    is that
 only    three   factors are used, namely, time, the historical number of publications, and the historical number of coauthors.
  Analyzing massive data to track the scientific careers of researchers would help to advance our understanding  of how   collaboration patterns   evolve.
The career stage of a researcher  is worth   considering as an influencing factor.
 It would be interesting   to input
the rank  of the institutions to which   researchers belong,   the number of affiliations of past coauthors, the academic age,  and  reputation of researchers.

 \section*{Acknowledgments} The author thanks   Professor Jinying Su in the
National University of Defense Technology for her helpful comments
and feedback. This work is supported by the   National Natural   Science Foundation of China (Grant No. 61773020) and  National Education Science Foundation of China (Grant No. DIA180383).

\section*{Appendix A: Evidence  to support the motivation}

%the largest number of the researchers with $m$ publications   at the year $y$ and with $l$ coauthors before $y$ for any possible $(m,l)$.

\begin{figure*}[h]
\centering
% Use the relevant command to insert your figure file.
% For example, with the graphicx package use
\includegraphics[height=2.8   in,width=4.6    in,angle=0]{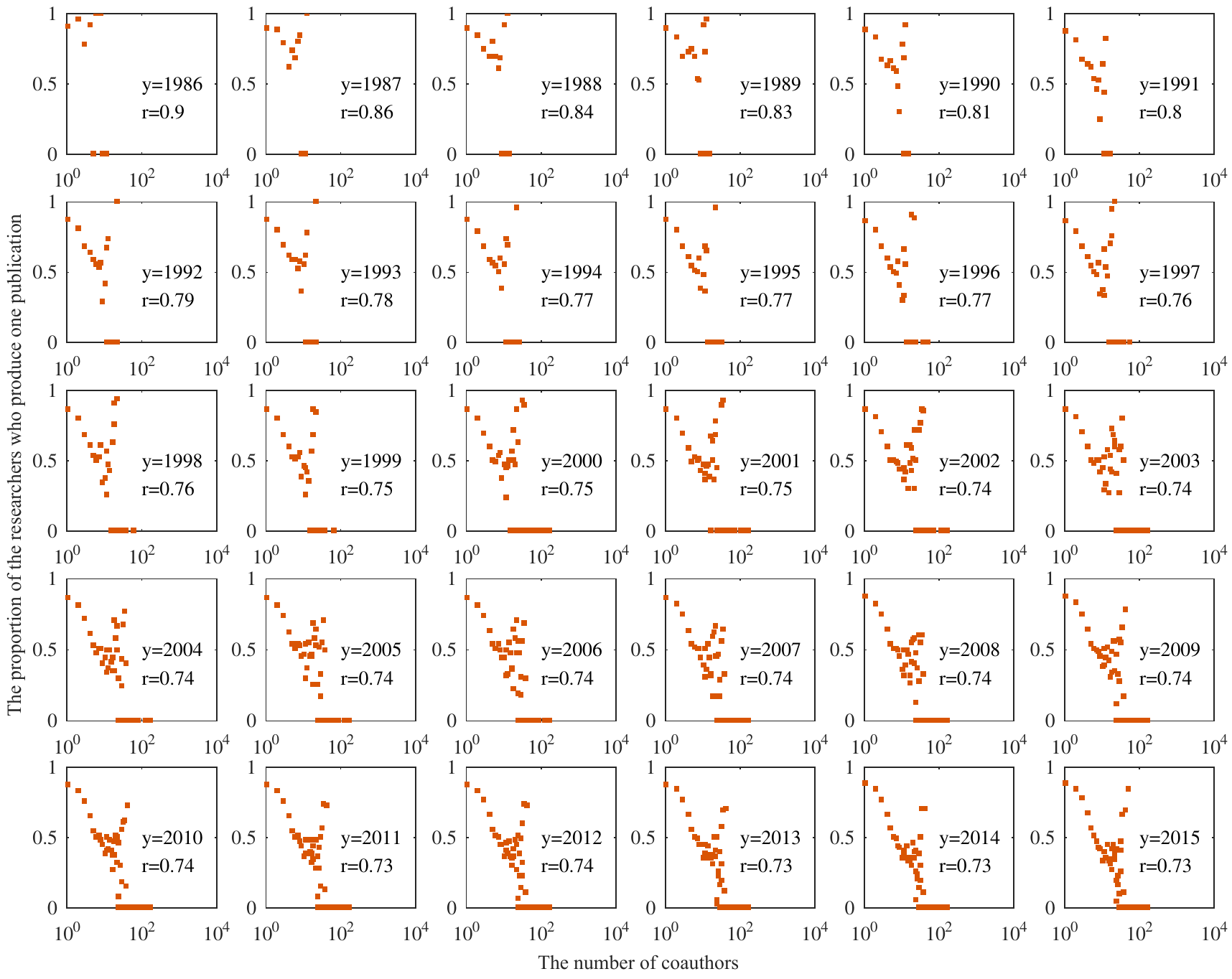}
% figure caption is below the figure
\caption{   {\bf The proportion of researchers who produced only one publication in the considered dataset.  }   The panels show that the proportion $r$  is more than 70\% in each year.   }
 \label{fig2e}
\end{figure*}

\begin{figure*}[h]
\centering
% Use the relevant command to insert your figure file.
% For example, with the graphicx package use
\includegraphics[height=2.1   in,width=3.3    in,angle=0]{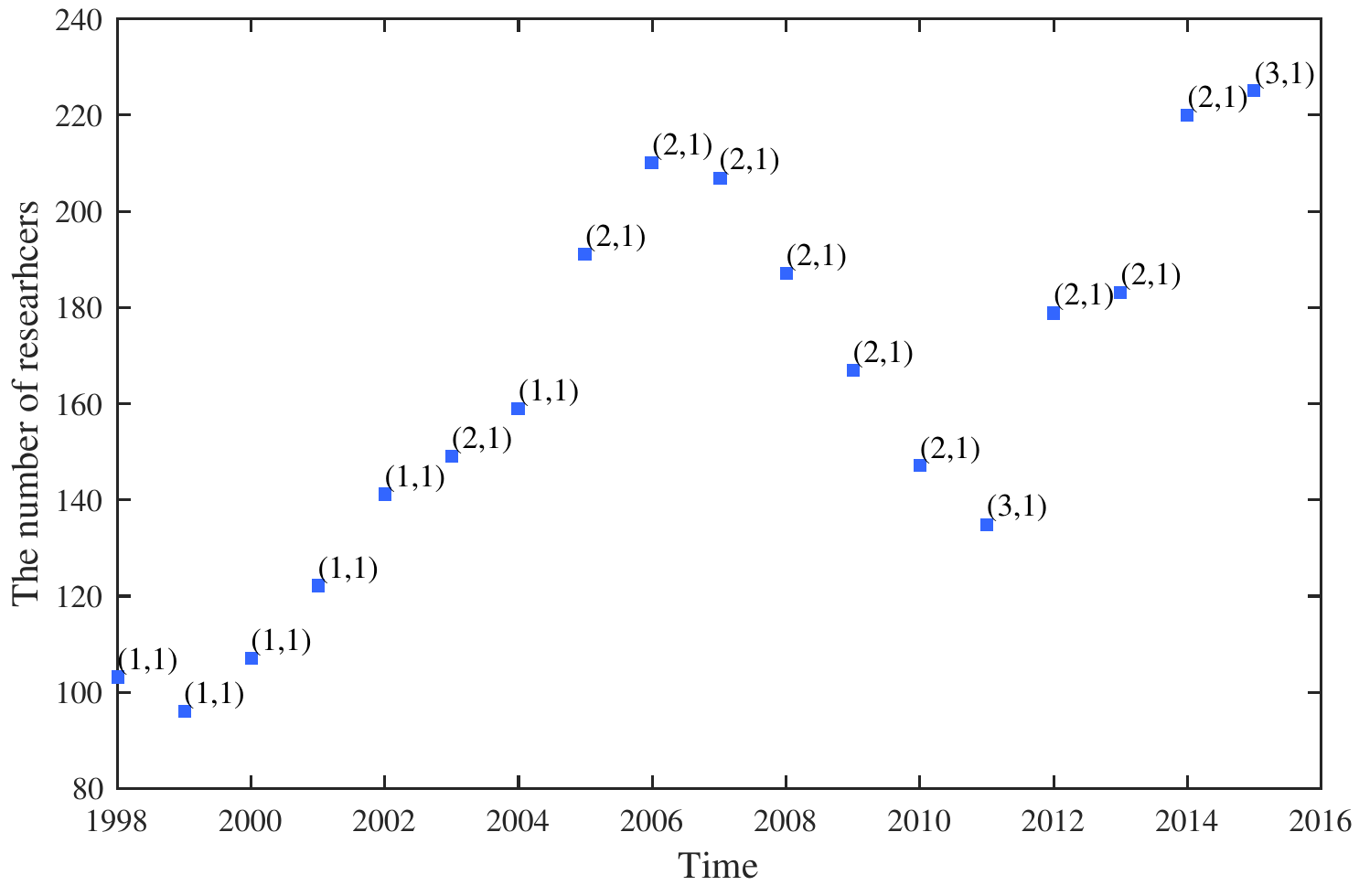}
% figure caption is below the figure
\caption{   {\bf The largest sample size of the KS test shown in Fig.~\ref{fig2} for  each year.  }  The panel shows the size of the largest  group of   researchers with $m$ publications  at year $y$ and with $l$ coauthors at $[1951,y-1]$ for any possible $(m,l)$.   }
 \label{fig2a}
\end{figure*}

\begin{figure*}[h]
\centering
% Use the relevant command to insert your figure file.
% For example, with the graphicx package use
\includegraphics[height=2.8   in,width=4.6       in,angle=0]{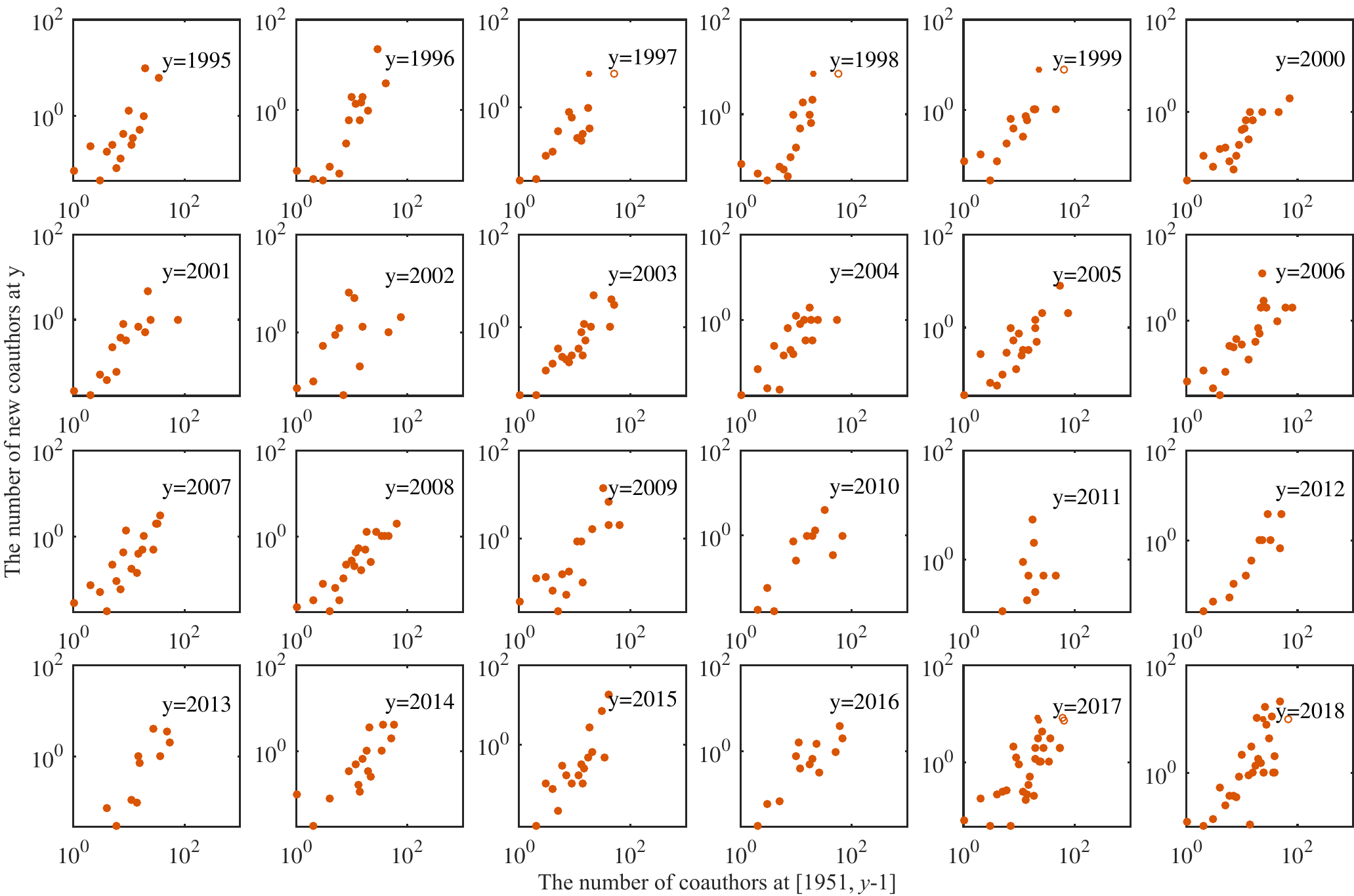}
% figure caption is below the figure
\caption{   {\bf  The cumulative advantage of attracting new coauthors. } The panels show  the average number of new coauthors
appearing in   year $y$ of  researchers whose    number of  coauthors at  $[1951, y-1]$  are the same.}
 \label{fig2d}
\end{figure*}

\begin{figure*}[h]
\centering
% Use the relevant command to insert your figure file.
% For example, with the graphicx package use
\includegraphics[height=2.8   in,width=4.6       in,angle=0]{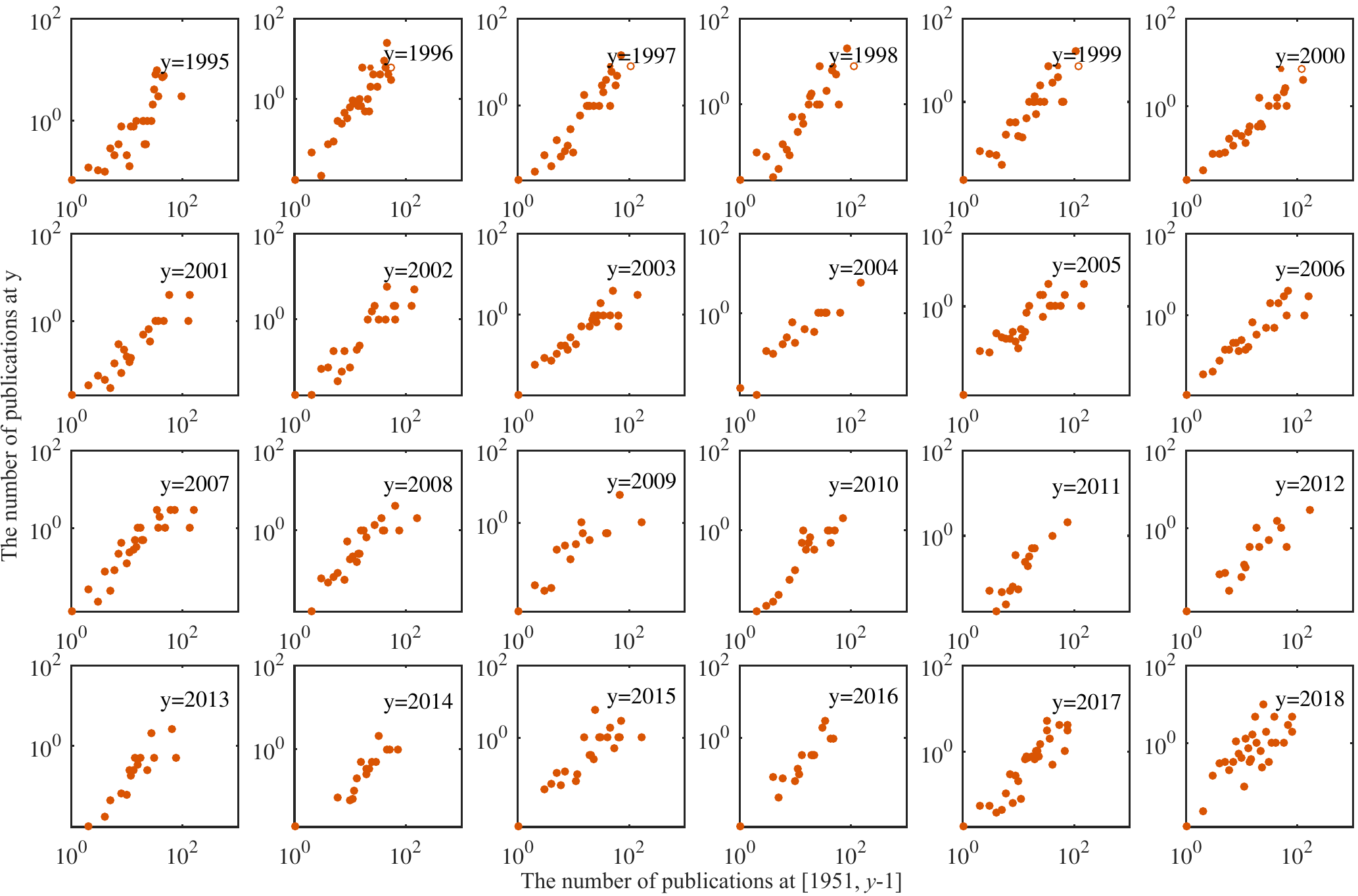}
% figure caption is below the figure
\caption{   {\bf  The cumulative advantage of producing publications. } The panels  show  the average number of   publications
  in  year $y$ of   researchers whose  number of  publications  at  $[1951, y-1]$ are the same. }
 \label{fig2c}
\end{figure*}

\begin{figure*}[h]
\centering
% Use the relevant command to insert your figure file.
% For example, with the graphicx package use
\includegraphics[height=2.8   in,width=4.6      in,angle=0]{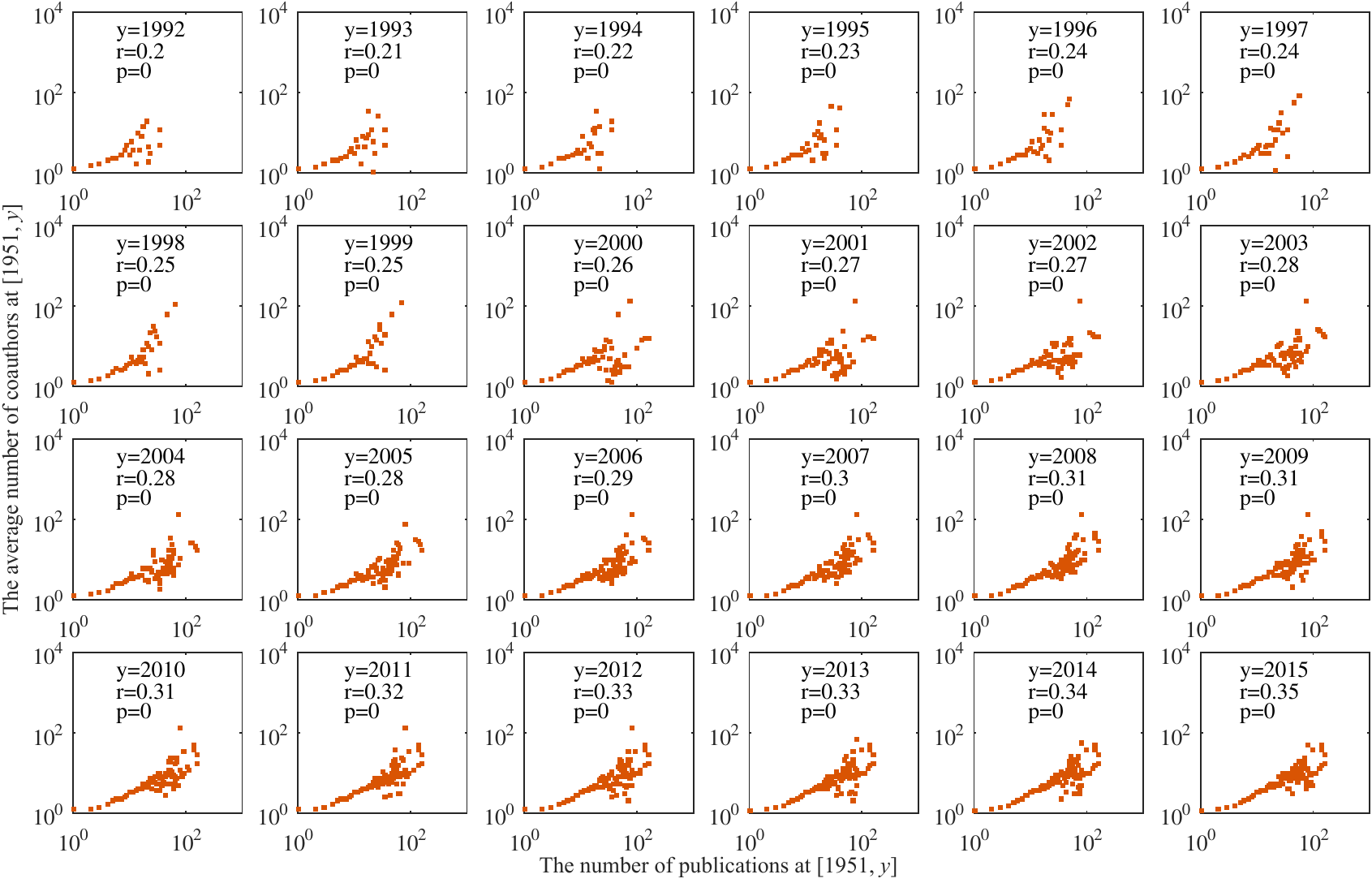}
% figure caption is below the figure
\caption{   {\bf The correlation
between the number of  publications   and  the number of coauthors. }  Consider the
researchers who produced publications at $[1951,y]$, where $y=1992,...,2005$.
The panels show   the average   number of coauthors
  at   $[1951,y]$ of   researchers who have  the same  number of  publications  at   $[1951,y]$.
  The Spearman correlation coefficient $r$ is significantly larger than $0$, $p$-value$<0.05$.}
 \label{fig2b}
\end{figure*}

Fig.~\ref{fig2e}
 shows    the proportion of researchers who  produced only one publication in the considered dataset.
The sample size  influences the $p$-value of  the KS test\cite{Moore-McCabe2014}: it can be    larger than $0.05$ in a large  sample and
 smaller than   $0.05$ in a small sample.
  Fig.~\ref{fig2a} shows that the sample sizes  of the tests shown in Fig.~\ref{fig2} are not very large.
  Fig.~\ref{fig2d}  shows      the cumulative advantage of attracting new coauthors
on researchers' historical number of coauthors.
Fig.~\ref{fig2c}
 shows      the cumulative advantage  of producing  new publications
on researchers' historical number of coauthors. Fig.~\ref{fig2b}   shows the positive correlation
between
the number of publications of
 a researcher  and  his or her number of coauthors.

\section*{Appendix B: The inapplicability of autoregressive models   }
 In statistics,   autoregressive models specify that the response variable depends linearly on its previous values with a stochastic term. The  advantage of those models is that they do not
  require much information;   only the self-variable series is needed.
 If the autocorrelation coefficients
 of the response variable series
 are   smaller than 0.5,  then autoregressive models are not suitable for the prediction task.
  The autocorrelation coefficient  of ${\boldsymbol y}=(y_1,...,y_T)$ with lag $l$   is defined  as
\begin{equation}
 r_l=    \frac{ \sum^{T-l}_{t=1}(y_t-\bar{\boldsymbol y})(y_{t+l}-\bar{\boldsymbol y})}  { \sum^{T}_{t=1}(y_t-\bar{\boldsymbol y})^2 },\label{eqA1}
\end{equation}
where $l<T$, and $\bar{\boldsymbol y} $ is the mean of ${\boldsymbol y}$'s elements\cite{Box1994}.

\begin{figure*}[h]
\centering
% Use the relevant command to insert your figure file.
% For example, with the graphicx package use
\includegraphics[height=2.4   in,width=4.6    in,angle=0]{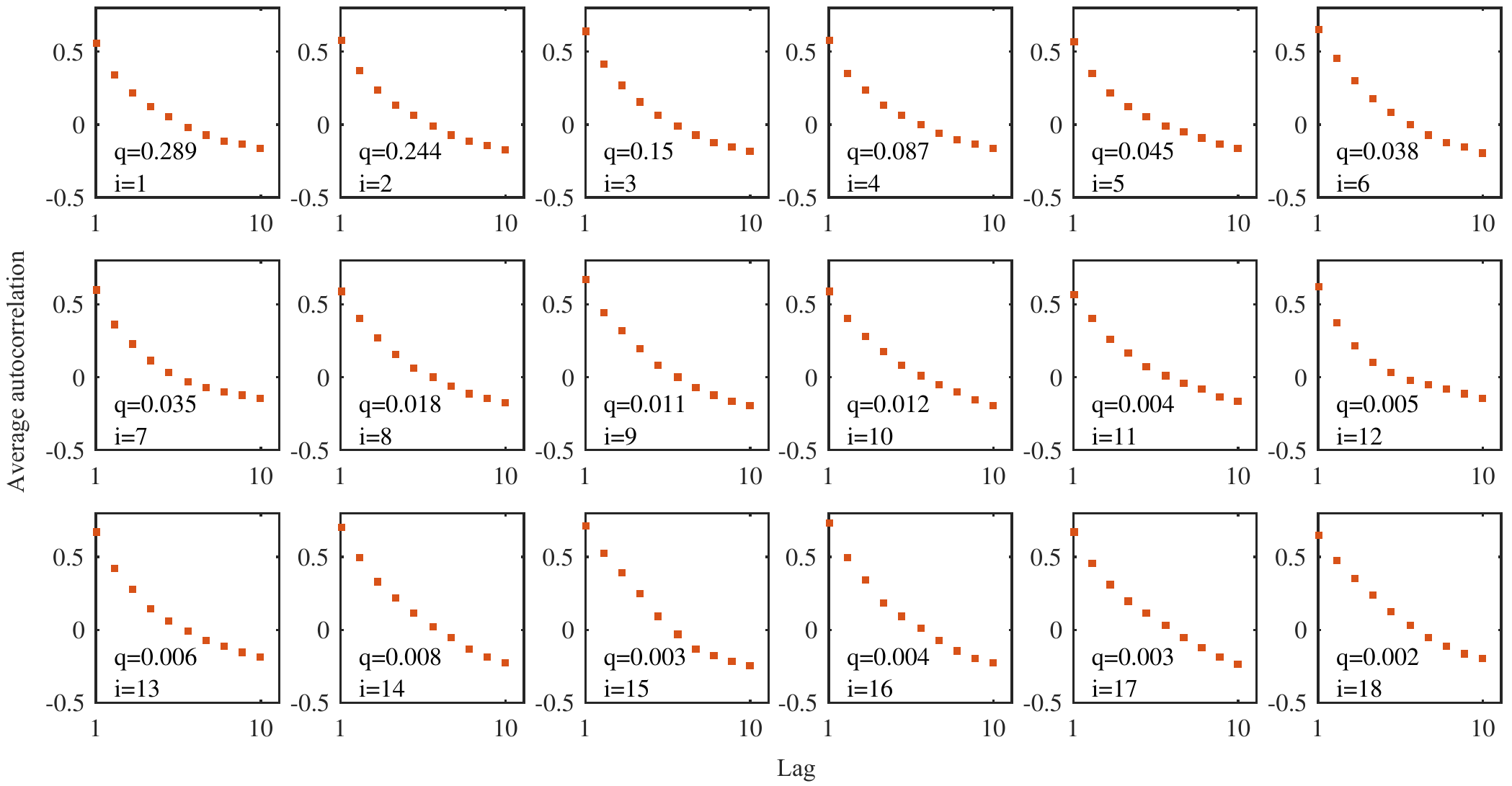}
% figure caption is below the figure
\caption{   {\bf     Autocorrelation coefficients
 of  the time series on the  cumulative number of coauthors.}  Let $k_s(t_j)$  be the  number of  coauthors of   researcher $s$
at   time interval $[1951, t_j]$.
Consider the  time series   ${k}_s=(k_s(t_X),...,k_s(t_{Z}))$, where
   $t_X=2000$,  and $t_{Z}=2018$.
   The panels show the average  autocorrelation coefficients of this time series  over the  group of    researchers in Set 4 who
have $i$ coauthors at $[1951,2000]$. Index $q$ is group proportion.}
 \label{figA1}
\end{figure*}

Consider a researcher $s$ in    test dataset Set 4. Consider
    his or her series of   the number of coauthors  ${\boldsymbol k}_s=(k_s(t_U),...,k_s(t_{V}))$, where
 $k_s(t_l)$  is his or her number of coauthors at $[T_{0},t_{l}]$  for     $t_U\leq t_l \leq t_V$.
 Substitute  the series   into Eq.~(\ref{eqA1}), and
  calculate  the autocorrelation coefficients.  Fig.~\ref{figA1} shows that   these coefficients with a lag~$>1$  are almost all smaller than 0.5. Therefore, an individual's   historical  number of coauthors  is    not sufficient to predict  his or her future  number of coauthors. This finding indicates that the autoregressive models  may not be suitable  for the prediction of the number of coauthors.

%Lags=$\{1,2,...,10\}$

\section*{Appendix C: An other example}

\begin{figure*}[h]
\centering
% Use the relevant command to insert your figure file.
% For example, with the graphicx package use
\includegraphics[height=3.    in,width=4.6    in,angle=0]{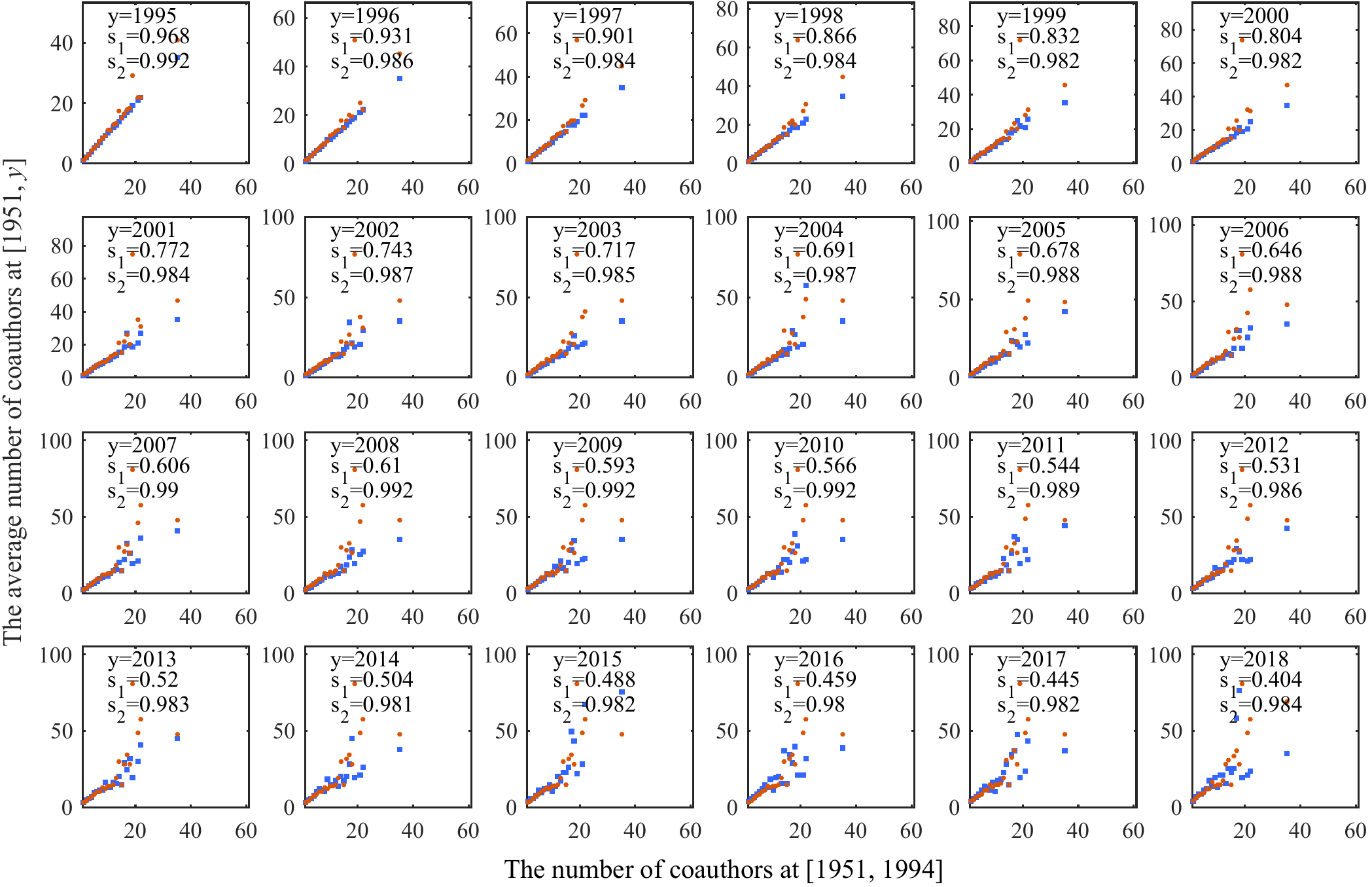}
% figure caption is below the figure
\caption{   {\bf  Fittings on the evolution of the number of coauthors for  researchers.} Consider the tested   researchers   who have  $k$ coauthors    at   $[1951,1994]$, where $k=1,...,60$.
The panels show
the average number of coauthors for  these researchers at    $[1951,y]$ ($n(k,y)$,  red dots) and the predicted number    ($m(k,y)$, blue lines).
  Index $s_1$ is the  Pearson  correlation coefficient
     calculated
 based on  the  list of researchers' number of  coauthors   and the  list of  their predicted number.
   Index $s_2$ is this coefficient based on the sorted
    lists.
}
 \label{fig7}
\end{figure*}

\begin{figure*}[h]
\centering
% Use the relevant command to insert your figure file.
% For example, with the graphicx package use
\includegraphics[height=3.   in,width=4.6     in,angle=0]{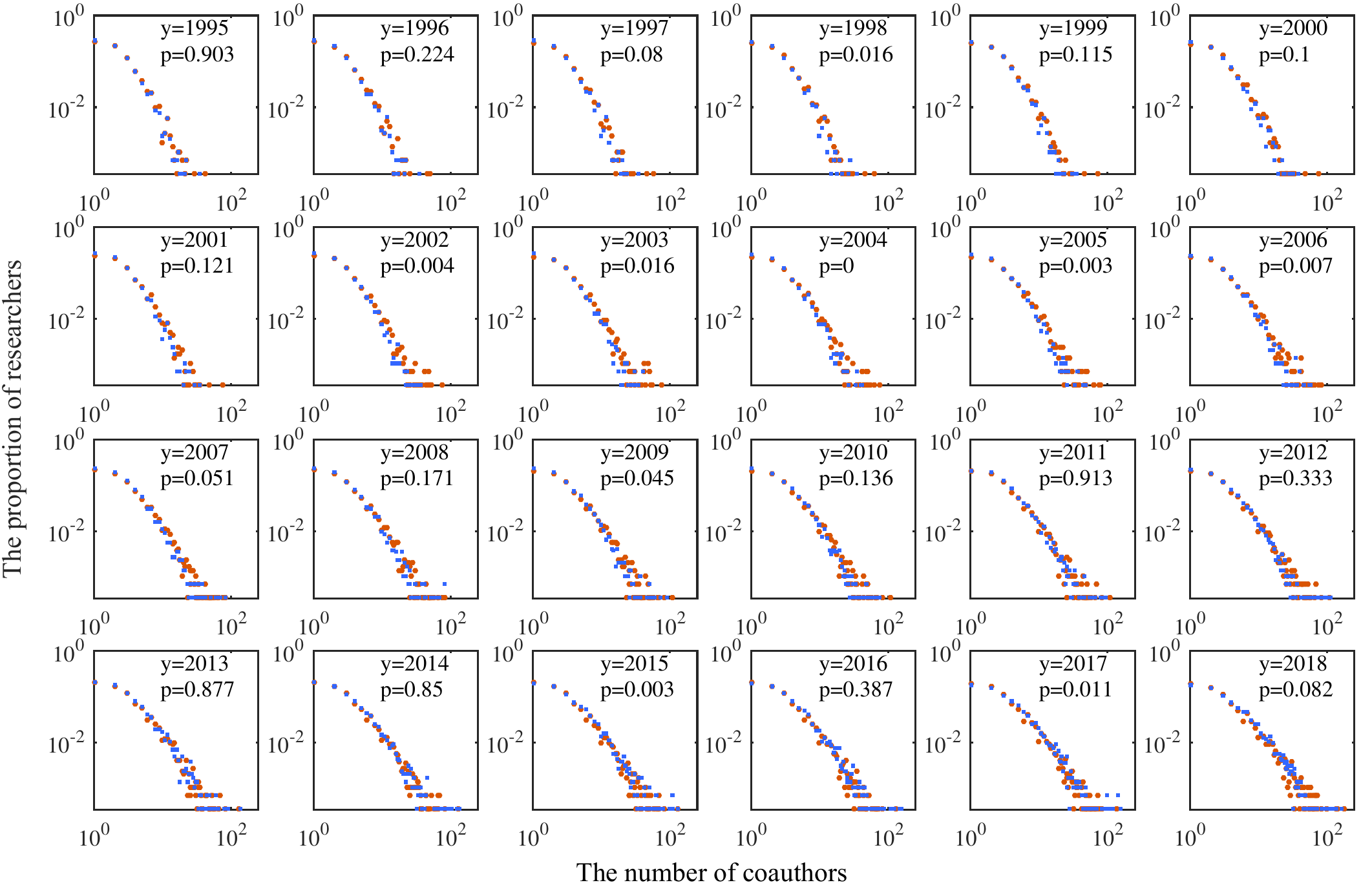}
% figure caption is below the figure
\caption{   {\bf  Fittings on  coauthor   distributions.}
 The panels show the   coauthor   distribution  of
 the tested  researchers
 at time interval   $[1951 ,y]$
    (red circles)  and the   predicted one
  (blue squares). When  $p>0.05$,   the KS test cannot reject   the   hypothesis that the compared distributions are  the same.  }
 \label{fig8}
\end{figure*}

\begin{figure*}[h]
\centering
% Use the relevant command to insert your figure file.
% For example, with the graphicx package use
\includegraphics[height=3.   in,width=4.6     in,angle=0]{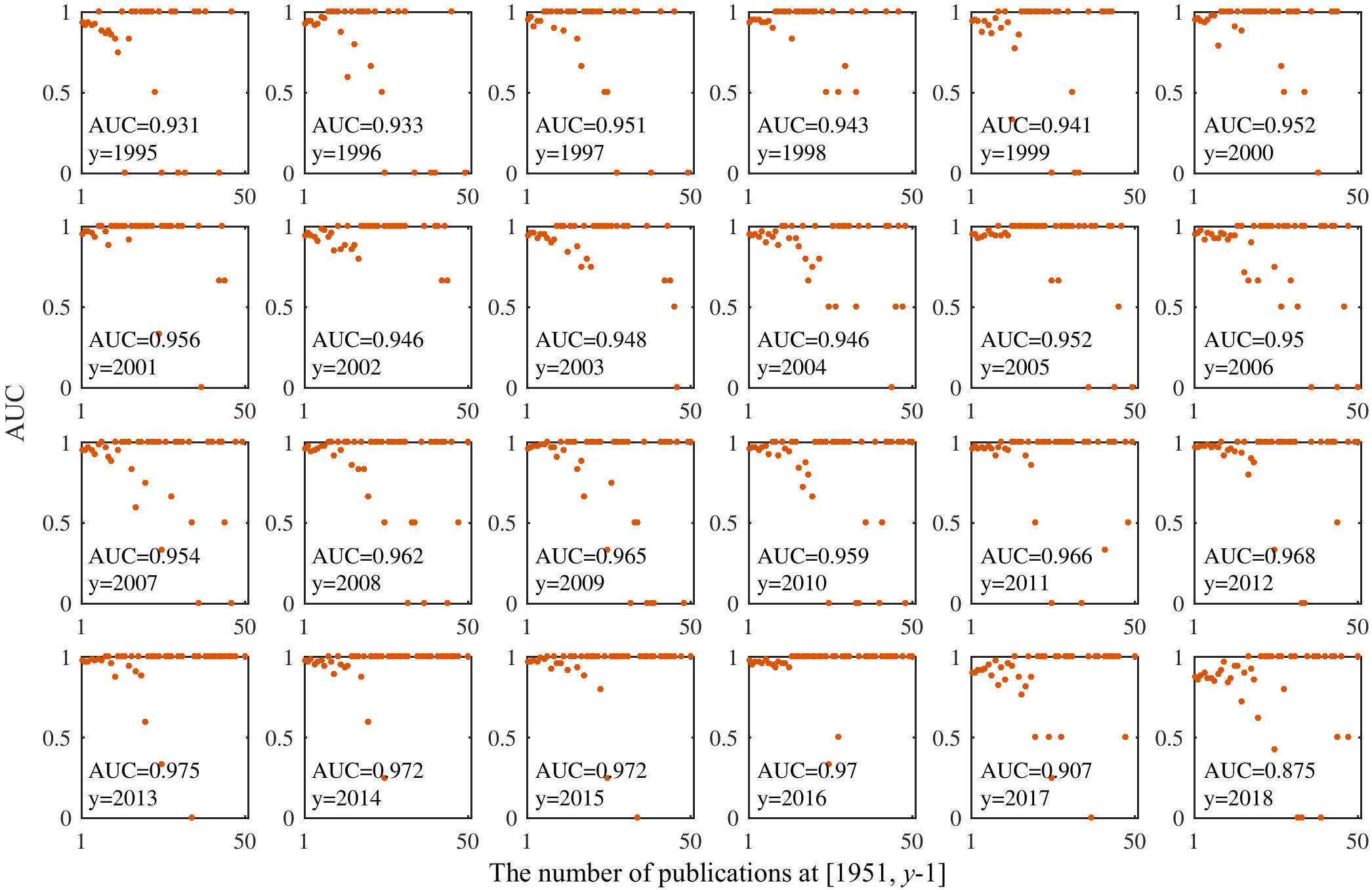}
% figure caption is below the figure
 \caption{        {\bf  The precision of  predicting  collaboration  events.}   The red dots show
   the   AUC    of  predicting the collaboration events  at     year $y$ for  the tested researchers   who produced  $i$ publications at $[1951,y-1]$, where $i=1,...,50$.  Index AUC is calculated based on all of the tested
researchers.           }
    \label{fig9}
\end{figure*}

The training and validation datasets here are Sets 5 and   6.
 The parameters  of the test dataset (Set 3)  are   $t_X=1994$, $t_Y=2010$, and
  $t_Z=2018$.  We   only predicted the publications for    99.98\% of the   researchers  in Set 3 who have no more than $60$ publications at the time interval $[T_0,t_X]$  and no more than $12$ annual publications at $[t_X,t_Z]$.
Figs.~\ref{fig7}-\ref{fig9} show
the prediction results of   our model.

%The validation dataset is Set 9.

\end{document}